\documentclass[aps,pra,reprint,showpacs,notitlepage,superscriptaddress,nofootinbib,twocolumn]{revtex4-2}
\usepackage{amssymb}
\usepackage{mathrsfs}
\usepackage{mathtools}
\usepackage{amsfonts}
\usepackage{graphicx}
\usepackage{amsmath}
\usepackage{dcolumn}
\usepackage{comment}
\usepackage{dsfont}
\usepackage[usenames,dvipsnames]{xcolor}
\usepackage[caption=false]{subfig}
\usepackage{booktabs}
\usepackage{epsfig}
\usepackage{soul}
\usepackage[normalem]{ulem}
\usepackage[colorlinks,linkcolor=blue,citecolor=blue,hyperindex,bookmarks=false,pdfstartview=FitH,urlcolor=blue]{hyperref}
\usepackage{appendix}
\usepackage{framed}
\usepackage{multirow}

\providecommand{\U}[1]{\protect\rule{.1in}{.1in}}
\definecolor{Blue}{rgb}{0.00,0.00,0.80}
\definecolor{Red}{rgb}{0.80,0.00,0.00}

\newcommand{\red}[1]{\textcolor{Red}{#1}}
\newcommand{\green}[1]{\textcolor[rgb]{0.00,0.70,0.50}{{#1}}}
\newcommand{\figpanel}[2]{\hyperref[#1]{\ref*{#1}(#2)}}

\newcommand{\avg}[1]{\langle #1 \rangle}
\newcommand{\occdiff}[0]{f(\varepsilon) - g(\varepsilon)}
\newcommand{\kB}[0]{k_\mathrm{B}}
\newcommand{\En}[0]{\varepsilon}

\newcommand{\reply}[1]{\textcolor{black}{#1}}
\newcommand{\correct}[1]{\textcolor{black}{#1}}

\begin{document}

\title{Optimizing energy conversion with nonthermal resources in steady-state quantum devices}

\author{Elsa Danielsson}
\affiliation{Department of Microtechnology and Nanoscience (MC2), Chalmers University of Technology, 412 96 Gothenburg, Sweden} 
\author{Henning Kirchberg}
\affiliation{Department of Microtechnology and Nanoscience (MC2), Chalmers University of Technology, 412 96 Gothenburg, Sweden}
\author{Janine Splettstoesser}
\affiliation{Department of Microtechnology and Nanoscience (MC2), Chalmers University of Technology, 412 96 Gothenburg, Sweden}

\date{\today}
\begin{abstract}
We provide a framework for optimizing energy conversion processes in coherent quantum conductors fed by nonthermal resources. Such nonthermal resources, which cannot be characterized by temperatures or electrochemical potentials, occur in small-scale systems that are smaller than their thermalization length. 
Using scattering theory in combination with a Lagrange multiplier method, we optimize the device's performance based on the efficiency, precision, or a trade-off between the two at a given output current. The transmission properties leading to this optimal  performance are identified. 
We showcase our findings with the example of a refrigerator exploiting experimentally relevant nonthermal resources, which could result from competing environments or from light irradiation. We show that the performance is improved compared to a device exploiting a thermal resource. 
Our results can serve as guidelines for the design of energy-conversion processes in future nanoelectronic devices.
\end{abstract}

\maketitle

\section{Introduction}\label{secIntro}

Conventional thermoelectric heat engines and refrigerators are energy converters that operate between two (or more) thermal reservoirs of different temperatures and electrochemical potentials. 
Thermoelectric energy conversion in nanoscale and mesoscopic conductors has recently attracted significant attention~\cite{Sothmann_2013,Haupt2013,QTDBook2018,Benenti2017Jun,Whitney2018May,Kirchberg2022,Campbell2025Apr}, driven by its versatile potential for efficient power production and cooling~\cite{Giazotto2006Mar}. Due to the controllable energy-filtering properties of the conductor, nanoelectronic devices can act as highly tunable and efficient thermoelectric energy harvesters \cite{Hicks1993May,Mahan1996Jul,Whitney2015}. This has also been demonstrated in recent experiments using quantum dots or molecular junctions~\cite{Reddy2007,Josefsson2018Oct,Thierschmann2015Oct,Roche2015Apr,Jaliel2019Sep,Gehring2021Apr}.

In these small-scale devices, mostly operating at low temperatures, the thermalization length can however easily exceed the system size~\cite{Benenti2017Jun}. The electronic distributions in the contacts are hence possibly \textit{nonthermal}. 
This means that they cannot be characterized by temperatures and electrochemical potentials. 
Nonthermal electronic distributions can arise from the interplay between competing environments~\cite{Konig2010Sep,Tesser2023}, from irradiation by light fields~\cite{Lloyd-Hughes2021Jul,Maiuri2020Jan,Uehlein2025Mar,Cherubim2019} or even due to an external driving of some nearby operated device. They can also be created by controlled mixing of different thermal distributions~\cite{Altimiras2010Nov,Kovrizhin2012Sep, leSueur2010Jul} or by correlations between system and bath due to their strong coupling~\cite{Aguilar2024}\reply{, and even in other types of non-electronic systems, such as in trapped ions~\cite{Rossnagel2014} or active matter like bacteria ~\cite{Krishnamurthy2016Dec, Wiese2024}}. \reply{In electronic systems,} nonthermal distributions can \reply{in fact} be experimentally measured and fully characterized, for example by state spectroscopy~\cite{Altimiras2010Oct}.
Nonthermal distributions can be exploited as resource for energy conversion processes, potentially offering even greater advantages than their thermal counterparts~\cite{Sanchez2019Nov, Hajiloo2020Oct,Ciliberto2020Nov,Deghi2020Jul}. However, bounds on the optimal performance of energy conversion processes mostly resort to the thermal properties of the resources, such as the fundamental Carnot bound for the efficiency or the Curzon-Ahlborn bound~\cite{Whitney2018May}. There already exist some guidelines for optimizing the performance of nanoscale energy conversion processes in nonequilibrium devices---primarily targeting  coupling to macroscopic thermal reservoirs~\cite{Whitney2014Apr,Whitney2015,Landi2025,Chrirou2025Jul,Ryu_2022}. However, corresponding directives remain lacking for energy conversion processes that specifically exploit the unique features of nonthermal resources.
\begin{figure}[b]
    \centering
    \includegraphics[width=\linewidth]{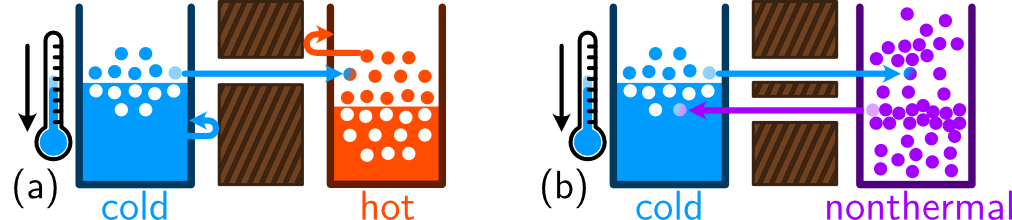}
    \caption{Sketch of the energy landscape of a two-terminal conductor with a central coherent scattering region. The goal is to extract heat from the left contact thereby cooling it. This is done by exploiting energy filtering adapted to the distribution of the right contact acting as a resource, which is either (a) hotter  than the left one or (b) given by a \textit{nonthermal} distribution. }
    \label{fig:Setup}
\end{figure}
We address this challenge by optimizing the performance of coherent electronic conductors with two contacts, where one of the contacts is characterized by an arbitrary nonthermal distribution, see panel (b) of Fig.~\ref{fig:Setup}\reply{, which is maintained like this in steady state}. Scattering theory is used to describe the coherent charge transmission between the contacts, which is experimentally achievable at the nanoscale~\cite{VanHouten1992,Brantut2012,Brantut2013,Jaliel2019Sep,Putzke2020,Kobayashi2021Sep,Balduque2025Apr}.
This means that our results are applicable to any type of coherent conductor where many-body interactions are weak enough that they can be treated within a mean-field approach and where all inelastic processes that cause dissipation or decoherence occur in the contacts~\cite{Christen_1996}. 

To quantify the performance of this generic nonthermal situation, we use general transport currents to define the input as well as the output of the conversion process and the resulting generalized efficiencies, since heat flows are often not a meaningful concept any longer. \reply{In fact, defining work in quantum systems interacting with different types of environments is often challenging~\cite{Talkner2020,Landi2021Sep,Silva2021Oct}. Specifically in quantum transport, work and heat are defined in terms of electrochemical potentials~\cite{Benenti2017Jun}, which are not defined for nonthermal distributions. A current that can instead be defined is the entropy current\reply{~\cite{Deghi2020Jul,Acciai2024Feb,Tesser2025}, the change of entropy in a terminal, which is found by accessing the electronic distributions of the terminals.}} As \correct{the} input current from the resource, \correct{the} entropy \correct{current is} a meaningful quantity; a resource is ``used up" when its entropy is increased, allowing by virtue of the second law that the entropy is decreased elsewhere in the device. 

In addition to generic output currents and the related efficiencies, we also optimize the precision of these generic output currents as well as trade-off relations between precision and efficiency in the spirit of recently studied thermodynamic uncertainty \correct{relations}~\cite{Barato2015Apr,Gingrich2016,Pietzonka2018May}. 

Concretely, our optimization aims at finding the ideal transmission properties of the conductor with respect to the mentioned performance quantifiers, provided a fixed desired output current. To treat this constraint of a given output current, we use a Lagrange multiplier method, extending the strategy of Refs.~\cite{Whitney2014Apr,Whitney2015,Chrirou2025Jul}. This further allows us to expand our optimization to additional tunable parameters, such as externally applied gate potentials.

 We find that the performance of the device is optimized by a series of boxcar-shaped energy filters. We identify crossing points between characteristic spectral quantities which set the positions and widths of these energy filters. This provides guidelines for the design of nanoelectronic devices for energy conversion processes with nonthermal resources.
By working out the ideal performance, we also provide 
 bounds for realistic devices. Consequently, this means that we identify the largest output current, the highest efficiency at given output current and the best precision that one can possibly achieve in a two-terminal coherent conductor \correct{exploiting} a given nonthermal resource. 

While our results hold for any steady-state energy-conversion task, we apply our findings to the experimentally relevant challenge of cooling, namely extracting heat from the cold thermal contact of the device, as indicated in Fig.~\ref{fig:Setup}. We investigate two realistic situations: one where the nonthermal distribution is the result of two competing thermal distributions; and one where a peak and a dip occur on top of a thermal distribution, which could arise from light irradiation at a given frequency. We find that by designing an appropriate series of boxcar-shaped energy filters, a higher cooling power can be obtained compared to the equivalent thermal resource. Furthermore, the precision and the efficiency of the conversion process exceed the ones that would be obtained with a thermal resource.
This underlines the importance of designing energy-converters while accounting for nonthermal effects.

The model of the two-terminal device with a nonthermal resource is provided in Sec.~\ref{sec:method}, together with the performance quantifiers of interest and how they can be evaluated from scattering theory. Then, the optimization procedure is presented in detail in Sec.~\ref{sec:opt}. In Sec.~\ref{sec:optimal_fridge}, we demonstrate our results with the example of cooling using two experimentally relevant example nonthermal distributions. Further technical details are given in Appendices~\ref{app:crossings}-\ref{app:numerical}.

\section{Model and Theoretical Approach}\label{sec:method}
This section introduces the considered model for the two-terminal coherent conductor with a nonthermal resource. Furthermore, all relevant performance quantifiers are defined and we show how to evaluate \correct{them} using scattering theory for quantum transport.

\subsection{Two-terminal conductor with a nonthermal contact}\label{sec:model}
The subject of this study is a nanoelectronic conductor operating coherently between two terminals, as sketched in Fig.~\ref{fig:Setup}. The two macroscopic contacts have very different characteristics. The left contact is described by a thermal distribution, where the occupation probability of particles at energy $\En$ is given by a Fermi function $f(\En)=\left(1+\exp[(\En-\mu)/\kB T]\right)^{-1}$, with temperature $T$ and electrochemical potential $\mu$ (set to reference throughout the paper as $\mu = 0$). 
By contrast, the distribution of the right contact can be \textit{nonthermal} (see panel (b) compared to panel (a)). This means that the particles' occupation probabilities $g(\En)$ \correct{are} generally \textit{not} given by a Fermi function, but only need to fulfill the constraint $0\leq g(\En)\leq 1$, due to the Pauli exclusion principle. Throughout this paper, we assume that both electron \correct{distributions} are given and constant, since the device is \correct{ kept in a} steady-state \correct{situation}.

The nonthermal contact is considered to be the resource for energy conversion in this manuscript (and compared to an equivalent hot thermal contact). 
Such \textit{steady-state} nonthermal distributions occur when the length scale of the device is smaller than its thermalization length~\cite{Benenti2017Jun} or the extraction times of particles are much shorter than the thermalization time. This happens for example in hot carrier solar cells~\cite{Konig2010Sep,Ahmed_2021,Tesser2023} or other devices where different environments compete or are mixed with each other~\cite{Altimiras2010Nov,Sanchez2019Nov,Altimiras2010Oct} or in systems steadily irradiated by coherent light~\cite{Lloyd-Hughes2021Jul,Maiuri2020Jan,Uehlein2025Mar,Cherubim2019,Sarkar_2022}, but nonthermal distributions can also be a result of energy filtering~\cite{Heremans2008Jul,Bahk2013Feb}. The act of energy filtering by the conductor, which we here exploit for cooling, can therefore in principle also affect the distributions in the terminals, which can be taken into account in self-consistent manner~\cite{Tesser2023,Bertin-Johannet2025Jul}. However, accounting for this would require knowing the specific circumstances under which the nonthermal distribution is created; here, we instead focus on generic nonthermal and thermal distributions in the terminals. Characteristic examples of such nonthermal distributions are discussed in Sec.~\ref{sec:optimal_fridge}. 

To utilize the resource to generate a useful output, the energy-dependent transmission properties of the central conductor are exploited.  
Many-body interactions in the coherent central conductor are assumed to be weak~\cite{Christen_1996}. Therefore, it is possible to capture the properties of the conductor by a transmission probability, $\mathcal{D}(\En) \in [0,1]$. In Fig.~\ref{fig:Setup}, the conductor acts as an energy filter \correct{with} $\mathcal{D}(\En) = 1$ where electrons are transmitted and $\mathcal{D}(\En) =0 $ where they are reflected. 

We assume a single-channel conductor, however the discussion in this paper can be straightforwardly extended to the multi-channel case. Then one would need to employ an \mbox{$(N+M)\times(N+M)$} scattering matrix, where $N$ and $M$ are the numbers of channels connecting the conductor to the left and right contacts.  The transmission probability is then replaced by $\mathcal{D}(E)=\mathrm{Tr}\{t^\dagger_\mathrm{LR}(E)t^{}_\mathrm{LR}(E)\}$, where $t_\mathrm{LR}$ is the transmission amplitude matrix (a sub-block of the scattering matrix) between left and right contact and the trace is taken over all channels.

The transmission function is the quantity that is optimized in Sec.~\ref{sec:opt}.  All other quantities and functions are assumed to be given, with the exception of a possible second variable in Sec.~\ref{sec:opt_two}. This perspective reflects designing a nanoelectronic conductor for a particular task and known constant electronic distributions. 

\subsection{Transport observables from scattering theory}\label{sec:scattering}
The device \correct{in} Fig.~\ref{fig:Setup} acts as an energy converter where energy is carried by electrons or holes in the Fermi sea. Heat transport due to phonons can typically be neglected in low-temperature experiments~\cite{Benenti2017Jun}. Note however that the nonthermal distributions treated in this manuscript \correct{could} indeed be the result of interactions between electrons and phonons or photons.

To analyze the performance of the two-terminal conductor as an energy converter, both the output of the conversion and the input provided by the nonthermal resource need to be characterized. Therefore, we calculate the particle, energy, and entropy currents \textit{out of} a contact flowing towards the scatterer (the coherent conductor). Assuming that interaction effects beyond the mean-field level can be neglected, scattering theory is used to calculate these transport quantities. \reply{The average current is derived from a current operator $\hat{I}^z(t)$ by taking the expectation value, yielding}~\cite{Moskalets2011Sep}
\begin{equation} \label{eq:generic_current}
    I^z = \int_{-\infty}^\infty d\varepsilon\, z(\varepsilon) \mathcal{D}(\varepsilon)(f(\varepsilon) -g(\varepsilon)).
\end{equation}
Here, $z(\En)$ is the current coefficient which describes the type of current in question. Some notable examples are $z(\En)=\pm 1/h$ for the particle current $I^{N,\mathrm{L/R}}$ out of contact L/R and $z(\En)=\pm(\En/h)$ for the energy current $I^{\En,\mathrm{L/R}}$ out of contact L/R. Further, $z(\En)=(\kB/ h) \ln[g(\En)/(1-g(\En)]$ gives the entropy current $I^{\Sigma,\mathrm{R}}$ out of the right contact and $z(\En)=-(\kB/ h)\ln[f(\En)/(1-f(\En))]$ the entropy current $I^{\Sigma,\mathrm{L}}$ out of the left contact~\cite{Deghi2020Jul,Acciai2024Feb}. 

These currents fulfill the thermodynamic laws~\cite{Benenti2017Jun, Topp2015, Nenciu2007}, which implies the following conservation laws or limits:
\begin{subequations}
\begin{eqnarray}
    I^{\En,\mathrm{L}} + I^{\En,\mathrm{R}} = 0, \\
    I^{N,\mathrm{L}} + I^{N,\mathrm{R}} = 0, \\
    -I^{\Sigma, \mathrm{L}} -I^{\Sigma, \mathrm{R}} \geq 0.\label{eq:second_law}
\end{eqnarray}
\end{subequations}
Importantly, particle and energy currents are conserved and therefore just change sign when considered in the different contacts. This is different for the entropy current, where $\kB\ln[\frac{g(\En)}{1-g(\En)}]$ and $-\kB\ln[\frac{f(\En)}{1-f(\En)}]$ can be different, or in other words, entropy can be produced, only being constrained from below by the second law. 
Since the left contact is always thermal, we furthermore have $-\kB\ln[\frac{f(\En)}{1-f(\En)}]=(\En-\mu)/T$. This means that the entropy current is directly proportional to the heat current, $T I^{\Sigma,\mathrm{L}}=I^{Q,\mathrm{L}}$, which is obtained~\cite{Moskalets2011Sep,Benenti2017Jun} by setting $z(\varepsilon)=(\varepsilon-\mu)/T$ in Eq.~\eqref{eq:generic_current}. By contrast, in the nonthermal contact the heat current is generally not well-defined, due to the absence of a consistent definition of temperature and electrochemical potential. 

 The average current, given in Eq.~\eqref{eq:generic_current}, can be decomposed as an energy integral over \correct{a spectral current}. The spectral \correct{current is} given by
 \begin{equation}
    i^z_\mathcal{D}(\En) = z(\En)\mathcal{D}(\En)(f(\En) -g(\En)) .\label{eq:spectral_current_full}
\end{equation}
It will later be useful to also consider the  full-transmission spectral current,
\begin{equation}\label{eq:spectral_current_1}
     i^z(\En)\equiv i^z_1(\En)= z(\En)(f(\En) -g(\En)) ,
 \end{equation}
which is given by setting $\mathcal{D}(\varepsilon)=1$ in Eq.~\eqref{eq:spectral_current_full}. Importantly, since interactions beyond the mean-field level are neglected and we are employing a single-particle picture, it is possible to analyze the spectral (energy-resolved) properties of these transport observables. We will make use of this later when optimizing the energy conversion in Secs.~\ref{sec:best_output} and~\ref{sec:gen_eff}, when we analyze transport contributions in small energy windows.

Since the \textit{precision} is a further performance quantifier of interest, in addition to the expectation values of the currents, their fluctuations will be considered. \reply{The fluctuations of a current $I^z$ is obtained from scattering theory by taking the expectation value of an auto-correlation function~\cite{Blanter2000Sep}
\begin{equation}
    \tilde{S}^z(t,\tau) = \avg{\left\{\delta \hat{I}^z(t+\tau),\delta \hat{I}^z(t)\right\}},
\end{equation}
where $\delta\hat{I}^z(t) = \hat{I}^z(t)- I^z$. The zero-frequency noise, which is directly related to the variance of the current, is then found by integrating the autocorrelator over the times, giving the explicit expression~\cite{Blanter2000Sep},}
\begin{eqnarray} \label{eq:noise_full}
    S^z &=& \int_{-\infty}^{\infty} d \varepsilon\, z^2(\varepsilon)\mathcal{D}(\varepsilon) (g(\varepsilon) g^-(\En) + f(\varepsilon) f^-(\En)) \nonumber \\ 
    &+& \int_{-\infty}^{\infty} d \varepsilon\, z^2(\varepsilon)\mathcal{D}(\varepsilon)(1-\mathcal{D}(\varepsilon))(g(\varepsilon) - f(\varepsilon))^2,    
\end{eqnarray} 
where $f^-(\En) = (1-f(\En))$ and $g^-(\En) = (1-g(\En))$.
Here, the first line represents thermal-like noise $S^z_{\mathrm{thermal}}$. Note however, that even though the first line has the functional form of the thermal noise, it involves the nonthermal distribution $g(\En)$. The second line is the shot noise $S^z_{\mathrm{shot}}$, also known as partition noise due to the factor $\mathcal{D}(\varepsilon)(1-\mathcal{D}(\varepsilon))$. It is non-zero only out of equilibrium, namely when $f(\En)$ and $g(\En)$ differ in some energy window.
Just as for the current, it will be useful to consider the full-transmission spectral noise that will be used when optimizing precision in Sec.~\ref{sec:gen_noise}.
It will turn out that only the full-transmission thermal noise contribution will be required,
\begin{eqnarray}\label{eq:spectral_noise}
        s^z(\En) &\equiv &s^z_{\mathrm{thermal}}(\varepsilon)  \nonumber \\  
        &=& z^2(\varepsilon)\mathcal{D}(\varepsilon)(g(\En)g^-(\En) +  f(\En)f^-(\En)),
\end{eqnarray}
where $S^z_{\mathrm{thermal}}=\int_{-\infty}^\infty d\varepsilon s^z_{\mathrm{thermal}}(\varepsilon)$.

\subsection{Performance quantifiers}\label{sec:efficiencies}

Our objective is to optimize the performance of energy converters. Therefore, a crucial first step is to identify meaningful performance quantifiers. 
The first important aspect for the performance is naturally the actual \textit{desired output}. This is quantified here as a generic current, $I^x$, see the definition in Eq.~\eqref{eq:generic_current}, with current factor $x(\varepsilon)$. 
For the matter of clarity and without loss of generality, we here always take the output current to be \textit{positive}, $I^x\geq 0$. Consequently\correct{,} optimization of the output current, which we present in Sec.~\ref{sec:best_output} always consists of \textit{maximizing} $I^x$. Concretely, this means that if the desired output is negative following the definition in Eq.~\eqref{eq:generic_current}, one needs to set $x(\varepsilon)\to-z(\varepsilon)$ for the optimization procedure described below. This ensures that the total and spectral currents convey the same physical meaning.

For later applications, presented in Sec.~\ref{sec:optimal_fridge}, we will focus on a device that acts as a refrigerator. This means that the cooling power, namely the heat current out of the (cold) thermal contact, $I^x\rightarrow I^{Q,\mathrm{R}}$ is the desired output of the energy conversion process. 

The next matter is how \textit{efficient} the generation of this output current is. We therefore also need to quantify the resource. Keeping the discussion as general as possible, we take as the input from the resource a generic current $I^y$ and define the efficiency as the fraction between output and input current
\begin{equation}\label{eq:def_efficiency}
    \eta = c_\eta\frac{I^x}{I^y} .
\end{equation}
Here, we also assume that the input current is always positive, such that the efficiency is a positive number. The factor $c_\eta$ is used to normalize the efficiency such that it is dimensionless. Usually, the efficiency is defined from the start as dimensionless, e.g., as the fraction between cooling power and input electrical power. However, the generic currents considered here could possibly have different dimensions. The constant $c_\eta$ is then found when constructing a meaningful efficiency based on the second law of thermodynamics. Since $c_\eta$ is merely a constant, it does not affect the following optimizations or results. 

The strategy of how to define a meaningful, dimensionless efficiency becomes clear from the concrete example considered in Sec.~\ref{sec:optimal_fridge}, where we study the cooling power in the left contact. Since the right, resource contact is not thermal, the resource is typically \textit{not} power or a heat current, as one would expect from a standard Peltier element or absorption refrigerator. Instead, motivated by the constraints put on energy conversion by the second law, we aim at characterizing the resource by an entropy current $I^y\rightarrow -I^{\Sigma,\mathrm{R}}$ (which is positive when entropy is produced in contact R). 
The constraints from the first law ensure energy and particle conservation, but \correct{cannot} yield insight into the heat transfer process. In the thermal (left) contact, the entropy reduction is proportional to the desired cooling power, namely to the heat current leaving the contact $I^x\rightarrow I^{Q, \mathrm{L}}=T I^{\Sigma, \mathrm{L}}$.  
The increase in entropy, $-I^{\Sigma,\mathrm{R}}>0$, implies that the right side acts as a resource (namely allowing by means of the second law that entropy is reduced elsewhere). Using the constraint by the second law~\eqref{eq:second_law}, we hence have $\eta=\frac{1}{T}\frac{I^{Q, \mathrm{L}}}{- I^{\Sigma,\mathrm{R}}}\leq 1$ with $c_\eta=1/T$.
Alternatives for quantifying the resource that have been considered previously could be free-energy currents~\cite{Hajiloo2020Oct,Manzano2020Dec,Tesser2023} (with respect to the temperature of the thermal contact), $I^y\rightarrow I^{\En,\mathrm{R}}-TI^{\Sigma,\mathrm{R}}$. 

Furthermore, it is important in nanoscale conductors that the output of the energy-conversion process is \textit{precise}. We hence aim at reducing the noise, where the performance quantifier is \correct{the noise $S^x$ of the current $I^x$} compared to the actual desired (average) output \correct{$I^x$}. This can be quantified \correct{by} the noise-to-signal ratio
\begin{equation}
    \mathrm{NSR} = c_\mathrm{NSR}\frac{S^x}{I^x}\correct{,}\label{eq:NSR_def}
\end{equation}
where $c_\mathrm{NSR}$ is again a normalization factor for the dimensionality. For charge currents, this ratio is known as the Fano factor~\cite{Blanter2000Sep}, where the constant takes the value  $c_\mathrm{NSR}=1/2e$.

It is well known from the output and the efficiency of standard engines that not both of these goals can be reached simultaneously, but that trade-offs need to be found. Here, we will consider one particular trade-off inspired by the thermodynamic uncertainty relation~(TUR)~\cite{Barato2015Apr,Pietzonka2018May}, which states that precision is bounded by entropy production. \reply{This TUR has been formulated in terms of transport currents~\cite{Brandner2018Mar,Kheradsoud2019Aug}, where it takes the form
\begin{equation}
    A_\mathrm{TUR}^{x, \sigma} = \frac{S^{x} \sigma }{(I^x)^2}\geq 2k_\mathrm{B},\label{eq:real_TUR_def}
\end{equation}
with $\sigma = -I^{\Sigma,L} - I^{\Sigma,R}$ being the total entropy production rate. Note that this bound is valid only for Markov processes and can be broken in quantum transport~\cite{Brandner2018Mar,Tesser2024May}, in particular in situations where the transmission probability is close to 1 in some energy intervals.}

\reply{We here consider a TUR-inspired factor
\begin{equation}\label{eq:TUR_def}
    A^{x,y} = c_A\frac{S^x I^y}{(I^x)^2}\correct{,}
\end{equation}
as a performance quantifier, where we again introduce a normalization factor $c_A$. 
Our definition of Eq.~\eqref{eq:TUR_def} is a more general trade-off factor, where $I^y$ is not specified from the start. Note that this factor is not bounded either---due to quantum effects and due to the fact that $I^y$ is a generic current and not necessarily the total entropy production rate. }

In the following, we will analyze how energy conversion using a nonthermal resource can be optimized concerning all the performance quantifiers in turn.

\section{Optimizing energy conversion}\label{sec:opt}

To optimize the conversion of an available energy resource into a useful output, such as cooling power, we use the previously introduced performance quantifiers as optimization objectives.
We have a general approach where the \textit{task} is to achieve a \textit{generic} output current, $I^x > 0$. The \textit {goal} is then to optimize different performance quantifiers, which are (A) the generic output currents themselves, (B) the efficiencies of their generation with \textit{generic} input currents (Eq.~\eqref{eq:def_efficiency}), (C) the precision of these output currents by the reduction of their noise (Eq.~\eqref{eq:NSR_def}), and finally (D) a combinations of these as a trade-off relation (Eq.~\eqref{eq:TUR_def}).
The aim of this paper is to find the optimal transmission function of the two-terminal quantum-scattering setup for the performance quantifiers. This approach, based on scattering theory, provides insights to guide the design of energy-conversion experiments and sets bounds on achievable performance, whenever independent transmission channels are present in  nanostructures~\cite{VanHouten1992,Benenti2017Jun}.

We begin with the straightforward optimization of the desired output current shown in Sec.~\ref{sec:best_output}. 
As a next step, we optimize the other performance quantifiers under the constraint of a \textit{given} output current. The strategies for finding the transmission functions for optimizing these different constrained performance quantifiers are analogous to each other and they are similar to variational principles using  Lagrange multipliers~\cite{Whitney2014Apr,Whitney2015}. 
In Refs.~\cite{Whitney2014Apr,Whitney2015}, the efficiency at given output power in a two-terminal device with thermal contacts was shown to be reached by having a boxcar-shaped transmission probability. 
Here, we extend this approach to systems with nonthermal distributions and optimize furthermore the efficiencies of generic input and output currents---including those where the current coefficient $x(\En)$ is \textit{nonlinear} in  energy such as in the entropy current---and of the  precision, see also Ref.~\cite{Landi2025}. 

We find that the ideal transmission function for a given performance quantifier, namely for a given goal, is a series of $M$ boxcar-shaped transmissions, 
\begin{equation}\label{eq:general_boxcar_series}
    \mathcal{D}(\varepsilon) = \sum_{i=0}^M \Theta(\varepsilon - \varepsilon_{2i})\Theta(\varepsilon_{2i+1} - \varepsilon).
\end{equation}
We show that the number $M$, the position and the width of these boxcar-shaped sections are identified by determining \textit{relevant crossing points} between energy-resolved characteristic functions related to the performance quantifier in question and between occupation probabilities, as will be demonstrated below.

The optimizations shown in this section are completely general in terms of the choice of distribution functions and currents. The \textit{examples} that we use here always show cooling power, in the form of entropy reduction, as output and entropy production as input; the example for the nonthermal distribution that we use is composed of a thermal distribution modified by a dip-peak structure that it could result from light irradiation, see Sec.~\ref{sec:irradiation}. 

As mentioned in Sec.~\ref{sec:efficiencies}, useful currents for the operation under consideration are assumed to be positive.
In the following, we also absorb any dimensionality factors into the currents for simplicity, such that input and output currents (or noise and currents) have the same units by construction.

\subsection{Best output current}\label{sec:best_output}
We start by optimizing the output current $I^x$, defined through Eq.~\eqref{eq:generic_current}, possibly with $x(\varepsilon)\to-z(\varepsilon)$ to ensure $I^x \geq 0$ as the desirable output current. Technically, this optimization step is the most straightforward. We start by noting that the integrand of the general current in Eq.~\eqref{eq:generic_current} consists of three factors: the current coefficient $x(\En)$, the difference in occupation functions $\occdiff$, and the transmission function $\mathcal{D}(\En)$. We assume that the difference of occupation functions, $\occdiff$, is given for each energy $\varepsilon$ and we want to maximize a given output current---hence we have a given $x(\En)$---by finding the optimal transmission function $\mathcal{D}(\En)$. 
\begin{figure}[tb]
    \centering
    \includegraphics[width=\linewidth]{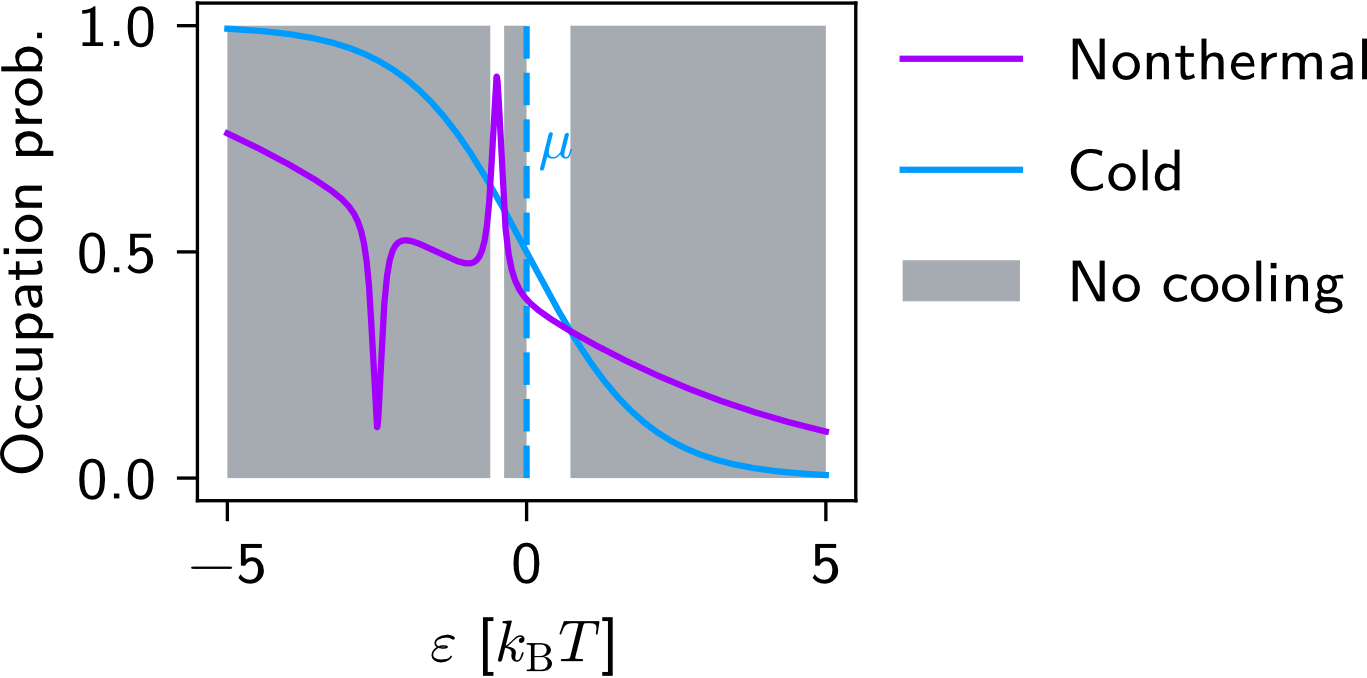}
    \caption{Occupation probabilities of the thermal contact of temperature $T$ (blue) and of the nonthermal contact (purple) as function of the energy $\mu\equiv0$ in the thermal contact. Energy windows in which the transmission should be set to zero in order to achieve optimal cooling power $I^x=I^{Q,\mathrm{L}}$ are shaded in gray.  }
    \label{fig:max_cool_example}
\end{figure}

The principle for the optimization is simple: in order to maximize $I^x$ one needs to allow for transport in those energy windows where the full-transmission spectral current $i^x(\En)$, defined in Eq.~\eqref{eq:spectral_current_1}, is positive and block transport whenever the full-transmission spectral current is negative. Sign changes in $i^x(\En)$ occur whenever the transport quantity $x(\En)$ changes sign or whenever the difference in occupation probabilities $\occdiff$ changes sign. This means that the output current is maximized  by setting $\mathcal{D}(\En) = 1$ for $x(\En)(\occdiff) > 0$ and $\mathcal{D}(\En) = 0$ for $x(\En)(\occdiff) < 0$.
The transmission function is thus a series of step functions,  
\begin{equation} \label{eq:opt_max_transmission}
    \mathcal{D}^{I}_{\mathrm{opt}}(\varepsilon) = \Theta(x(\En)(\occdiff)),
\end{equation}
determined by the crossing points between $x(\En)$ and 0, and between $f(\En)$ and $g(\En)$. These crossing points determine the parameters $\varepsilon_{2i},\varepsilon_{2i+1}$ entering Eq.~\eqref{eq:general_boxcar_series}. The result of this reasoning is shown for the example of a heat current out of the left contact (cooling power), such that $I^x=I^{Q,\mathrm{L}}$, in Fig.~\ref{fig:max_cool_example}. The regions in which one needs to have $\mathcal{D}(\En) = 1$ for optimal cooling are shown in white and the regions where  we choose $\mathcal{D}(\En) = 0$ to achieve optimal cooling are shown in gray.

\subsection{Best efficiency at fixed output current}\label{sec:gen_eff}
As a next step, we optimize the efficiency at a \textit{given} output current. We therefore follow the strategy of Ref.~\cite{Whitney2014Apr,Whitney2015,Chrirou2025Jul}, where it was shown that the transmission probability of a coherent, non-interacting two-terminal conductor with thermal contacts which maximizes the efficiency for a given power has a boxcar shape.
However, we have already seen in Eq.~\eqref{eq:opt_max_transmission} and in  Fig.~\ref{fig:max_cool_example} for the optimization of the output current that the nonmonotonicity of the occupation probabilities and hence of the spectral currents leads to a more intricate shape of the optimal transmission probability.  

\begin{figure}[tb]
    \centering
    \includegraphics[width=0.9\linewidth]{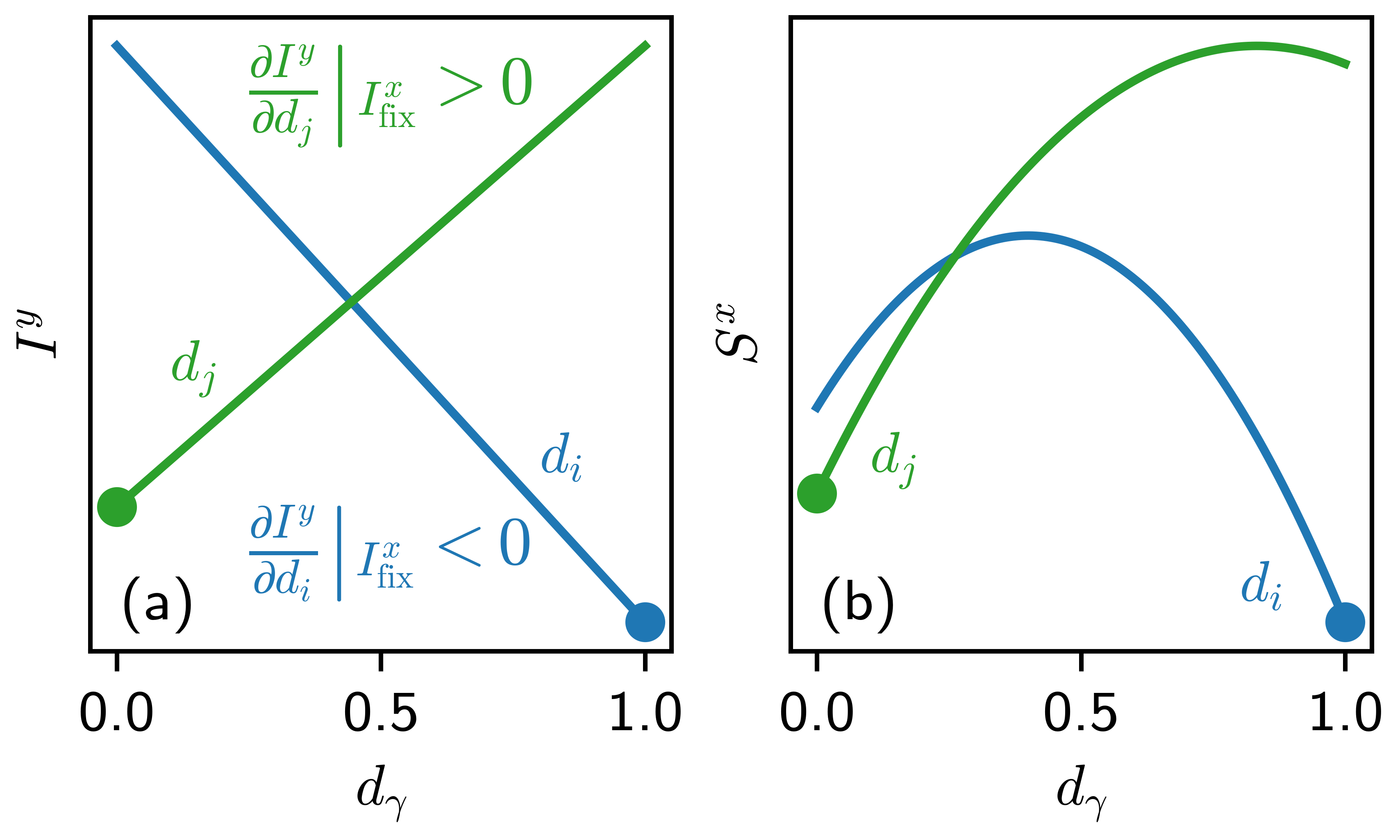}
    \caption{Illustration of how (a) currents change with transmission probability $d_\gamma$ (with $\gamma=i,j$) of two example energy slices $d_i$,$d_{j}$ and (b) how the noise changes with transmission probability of two example energy slices $d_i$,$d_{j}$. The minimum is in both cases always located at the boundary of the functions (indicated by filled circles).}
    \label{fig:linear_illus}
\end{figure}

The goal is to optimize the efficiency, Eq.~\eqref{eq:def_efficiency}, at a fixed output current $I^x$. 
As mentioned previously, since we neglect electron-electron interactions beyond mean-field level, we can consider the currents through small energy windows independently. 
We therefore now divide the energy integrals into small slices of width $\delta\En$, labeled by $\gamma$, in which we take the transmission probability as a variable $d_\gamma \in [0,1]$. The currents in each of these slices are hence
\begin{subequations}
\begin{eqnarray}
    I^x_\gamma(d_\gamma) &= \delta\En d_\gamma x_\gamma (f_{\gamma} -g_{\gamma}),\\
    I^y_\gamma(d_\gamma) &= \delta\En  d_\gamma y_\gamma (f_{\gamma} -g_{\gamma}).
\end{eqnarray}               
\end{subequations}
Here, we indicate by the subscript $\gamma$ the constant value that the functions $x,y,g,f$ take in the $\gamma$th energy slice. In the limit $\delta \En\to0$, the sum over all slices converges to the \correct{integral} given in Eq.~\eqref{eq:generic_current}. The task is now to solve the following  constrained problem:  find $\{d_\gamma\}$ such that $I^y = \sum_\gamma I^y_\gamma(d_\gamma)$ is minimized
with the constraint that $I^x = \sum_\gamma I^x_\gamma (d_\gamma)$ is fixed to $I^x=I^x_\mathrm{fix}$. In the spirit of the Lagrange-multiplier method, the constraint is now added to the input current to be minimized,
\begin{equation} \label{eq:discrete_setup}
    I^y = \sum_\gamma I^y_\gamma(d_\gamma) +\lambda \big(I^x_\mathrm{fix} - \sum_\gamma I^x_\gamma (d_\gamma)\big),
\end{equation}
with the Lagrange multiplier $\lambda$. 
The variation of the current $I^y$ with $d_\gamma$ under the constraint is
\begin{equation}\label{eq:discrete_cond}
    \left.\frac{\partial I^y}{\partial d_\gamma}\right|_{I^x_\mathrm{fix}} = \delta \En (y_\gamma-\lambda x_\gamma) (f_\gamma -g_\gamma).
\end{equation}
The right hand side is non-zero almost everywhere: it can be either positive or negative, meaning that the current increases or decreases linearly with a change in $d_\gamma$.
Crucially, since the current is linear in $d_\gamma$ (and hence not a concave function) there is no local minimum, see Fig.~\ref{fig:linear_illus}(a), and the problem can hence not be treated as a \textit{conventional} variational problem. 

Instead, we find a global minimum by considering the domain boundaries of $d_\gamma$. We therefore now \correct{ask how} the integrand behaves with variations of the transmission. If it is beneficial for the transmission function to increase for a certain energy, then the global minimum must be obtained when the transmission is equal to one for that point (and equal to zero in the opposite case). 
Explicitly, we set 
\begin{subequations}\label{eq:discrete_set_d}
\begin{eqnarray} 
    \left.\frac{\partial I^y}{\partial d_\gamma}\right|_{I^x_\mathrm{fix}} < 0 \longrightarrow d_\gamma = 1, \\
    \left.\frac{\partial I^y}{\partial d_\gamma}\right|_{I^x_\mathrm{fix}} > 0 \longrightarrow d_\gamma = 0, 
\end{eqnarray}    
\end{subequations}
 where examples for the first and second case are shown in blue and green respectively in Fig.~\ref{fig:linear_illus}(a). At the same time, the constraint of fixed output current must be fulfilled. This means that all transmissions should take the optimal values as in the equation above, while making sure that 
\begin{equation}\label{eq:constraint_discrete}
    I^x_\mathrm{fix} - \sum_\gamma I^x_\gamma (d_\gamma) = 0. 
\end{equation}
Since $d_\gamma$ also depends on the Lagrange parameter $\lambda$, this practically means finding $\lambda$ such that Eq.~\eqref{eq:constraint_discrete} is fulfilled, while letting each slice have their optimal value according to Eq.~\eqref{eq:discrete_set_d}. For discrete slices, it might not be possible to let every transmission value be at the optimal point and have the correct fixed current simultaneously. It is necessary to take the continuous limit.

\begin{figure}[bt]
    \centering
\includegraphics[width=\linewidth]{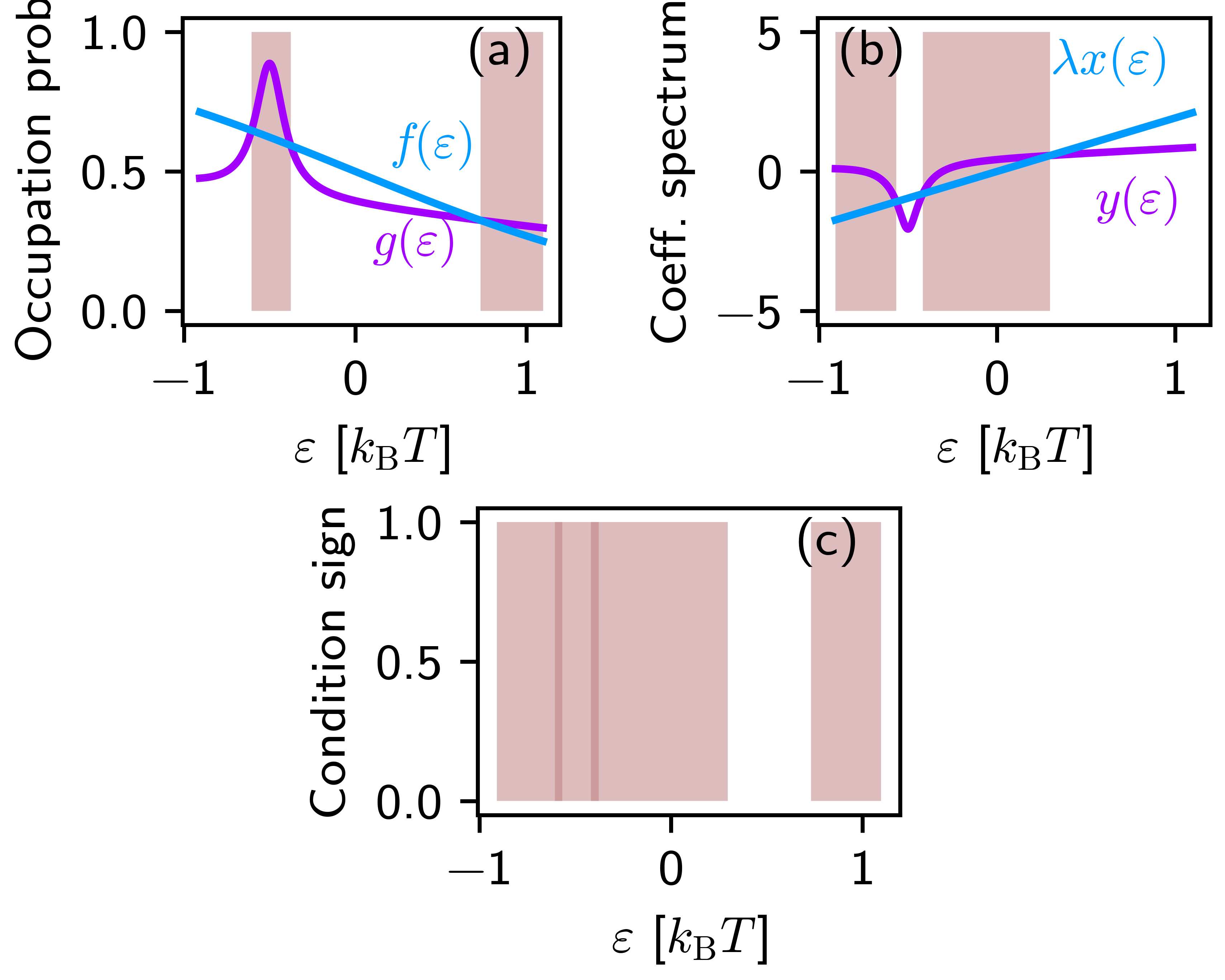}
    \caption{Finding the transmission function with crossing points. Purple lines are related to quantities of the nonthermal contact, blue lines to quantities of the cold one. (a) Two distribution functions, where red indicates regions where the second factor in the argument of Eq.~\eqref{eq:opt_eff_transmission} is positive. (b) The two current coefficients entering the first factor of Eq.~\eqref{eq:opt_eff_transmission}; $y(\En)$ gives the entropy current into the nonthermal bath and $\lambda x(\En)$ the entropy current out of the cold bath, weighted by the Lagrange multiplier. Panel (c) illustrates where in the energy spectrum the two factors are negative. In the fainter red areas, the step-function in Eq.~\eqref{eq:opt_eff_transmission} yields zero. In the white and in the darker red area, the factors in the argument of Eq.~\eqref{eq:opt_eff_transmission} are both positive or both negative, such that the transmission function is set to 1.}
    \label{fig:crossing_eff}
\end{figure}

We now take the continuous limit of the energy spectrum by letting $\delta \En \to 0$ to find a condition for the entire transmission function $\mathcal{D}(\En)$. The continuous equivalent of Eq.~\eqref{eq:discrete_setup} is 
\begin{eqnarray}
    I^y  &=& \lambda I^x_\mathrm{fix} + \int d\varepsilon \mathcal{D}(\varepsilon)(y(\varepsilon)- \lambda x(\varepsilon))(\occdiff)\nonumber \\&=& \lambda I^x_\mathrm{fix} + \int d\En (i^y_\mathcal{D} - \lambda i^x_\mathcal{D}) .    
\end{eqnarray}
In analogy to the discrete case, we aim to  extract the optimal transmission from  minimizing $I^y$. Following usual calculus of variations~\cite{Arfken}, the current under variation of the transmission function and constraint $I^x_\mathrm{fix}$ reads,
\begin{equation}
    \delta I^y|_{I^x_\mathrm{fix}} = \int d\En \frac{\partial}{\partial \mathcal{D}}(i^y_\mathcal{D}(\En) - \lambda i^x_\mathcal{D}(\En))\delta \mathcal{D}(\En) .
\end{equation}
Here, variational calculus reaches an impasse; as previously argued there should not exist any local minimum with respect to $\mathcal{D}(\En)$, meaning that setting $\delta I^y|_{I^x_\mathrm{fix}} = 0$ and solving like a regular variational problem yields nonsensical results. Instead, the continuous notion of a variation is combined with the knowledge from the discrete optimization by introducing an \textit{inequaltity}: $\delta I^y|_{I^x_\mathrm{fix}} < 0$. Furthermore, Eq.~\eqref{eq:discrete_cond} implies that the energy-resolved inequality lets us find the optimal transmission, by saturating the bounds of $\mathcal{D}(\En)$ to either 0 or 1 depending on the sign. Similar to how one in caluculus of variations solves for the stationary point with the integrand, we find
\begin{eqnarray} \label{eq:eff_cond}
    \frac{\partial}{\partial \mathcal{D}}(i^y_\mathcal{D}(\En) - \lambda i^x_\mathcal{D}(\En))\delta \mathcal{D}(\En) < 0   \nonumber \\
    \implies (i^y(\En) - \lambda i^x(\En))\delta \mathcal{D}(\En)  < 0.
\end{eqnarray}
Instead of directly allowing a solution for $\mathcal{D}(\En)$, this inequality gives the full transmission condition: it implies that for increasing $\delta\mathcal{D}(\En)>0$, it is beneficial to have $(i^y(\En) - \lambda i^x(\En)) < 0$ and for decreasing $\delta \mathcal{D}(\En) < 0 $ the opposite is true, $(i^y(\En) - \lambda i^x(\En)) > 0$. 
The condition can be simplified to setting $\mathcal{D}(\En) = 1$ wherever $(i^y(\En) - \lambda i^x(\En)) < 0$ and $\mathcal{D}(\En) = 0$ otherwise, following the logic that the transmission bounds should be saturated. 
The transmission function maximizing the efficiency at fixed output current is hence given by
\begin{equation} \label{eq:opt_eff_transmission}
    \mathcal{D}^{\eta}_{\mathrm{opt}}(\varepsilon) = \Theta((\lambda x(\varepsilon) -y(\varepsilon) )(\occdiff)) ,
\end{equation}
where $\lambda$ is fixed by the constraint $I^x_\mathrm{fix} = \int \mathcal{D}(\varepsilon)i^x(\En) d\En$. This strategy to find the optimal $\mathcal{D}(\varepsilon)$ is illustrated by an example in Fig.~\ref{fig:crossing_eff}.

We find an insightful interpretation of $\lambda$ in terms of a characteristic efficiency when inspecting the condition \correct{in Eq.}~\eqref{eq:eff_cond} further. It is readily rewritten as
\begin{equation}
     y(\varepsilon)(\occdiff) < \lambda x(\varepsilon)(\occdiff)
\end{equation}
for positive $\delta \mathcal{D}(\En)$. This is conveniently expressed with the full-transmission spectral currents of Eq.~\eqref{eq:spectral_current_1},
\begin{equation} \label{eq:spectral_inequality}
    i^y(\En) < \lambda i^x(\En).
\end{equation}

If both sides are divided by $i^y(\En)$ the condition is split into two cases, depending on if $i^y(\En)$ is positive or negative. 
Taking $\lambda$ to be positive, see Appendix~\ref{app:spectral}, the transmission function of Eq.~\eqref{eq:opt_eff_transmission} is then given by
\begin{eqnarray} \label{eq:two_contributions_D}
    \mathcal{D}^{\eta}_{\mathrm{opt}}(\varepsilon) &=& \Theta(i^y(\En)) \Theta\left(\frac{i^x(\En)}{i^y(\En)} - \frac1\lambda\right) \nonumber \\ &+&
    \Theta(-i^y(\En)) \Theta\left(\frac1\lambda - \frac{i^x(\En)}{i^y(\En)} \right) .
\end{eqnarray}
\begin{figure}[tb]
    \centering
    \includegraphics[width=\linewidth]{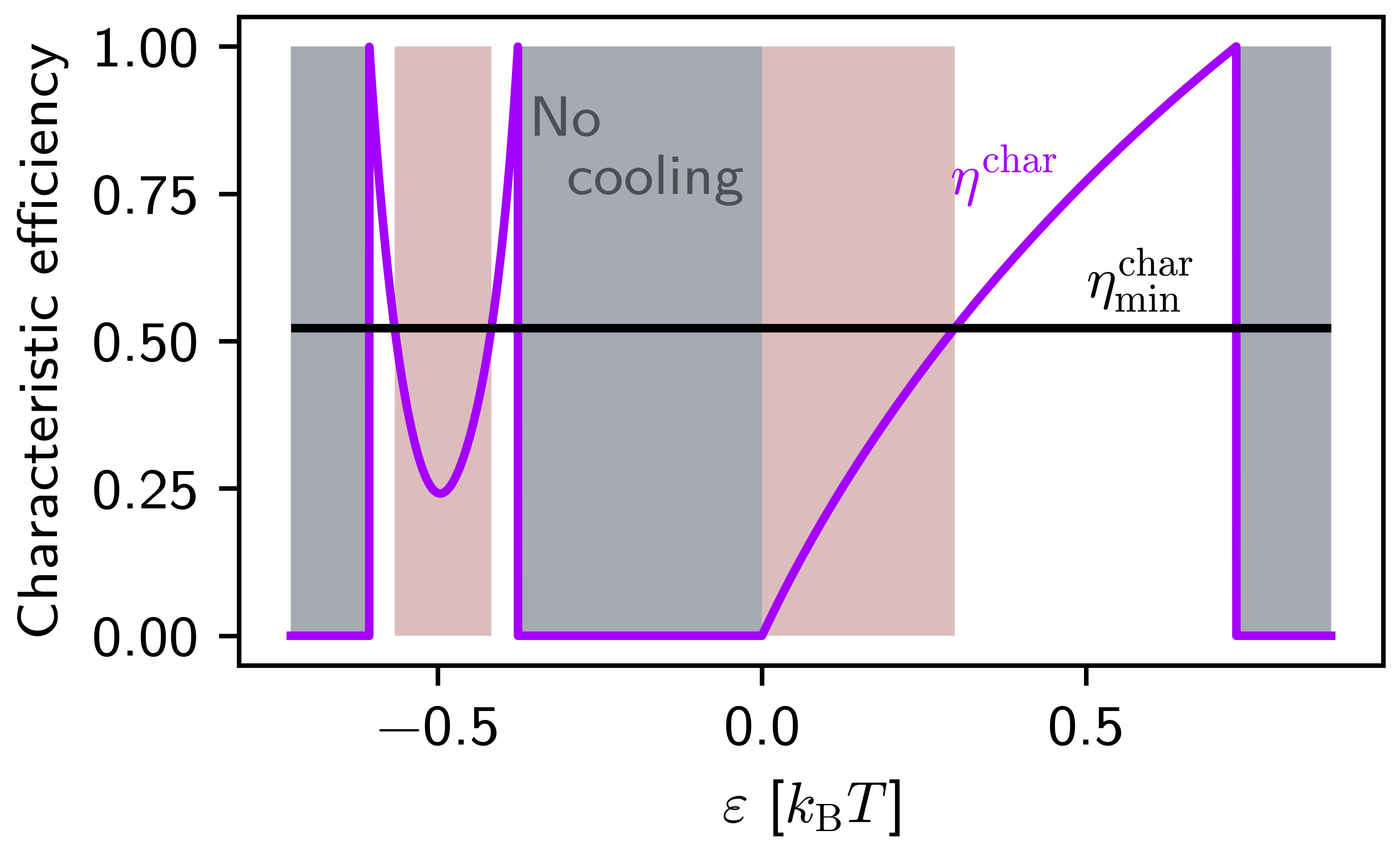}
    \caption{Example for the characteristic efficiency as function of energy (purple) compared to the minimum characteristic energy (black) fixed by the \textit{given} output current. When the spectral currents $i^x,i^y$ have opposite signs, no cooling takes place (gray areas). When $\eta^\mathrm{char}<\eta^\mathrm{char}_\mathrm{min}=1/\lambda$, cooling is relatively inefficient and hence excluded by the optimization. The white areas are the ones where $\mathcal{D}(\varepsilon)=1$ leads to optimal efficiency at a given output current.}
    \label{fig:char_eff}
\end{figure}
This expression can be simplified by classifying the spectral currents according to their sign in relation to each other. The first line in Eq.~\eqref{eq:two_contributions_D} only contributes wherever $i^x(\En), i^y(\En) >0$, which corresponds to the output being generated and resource spent, for a particular energy channel. The second line contributes when resource is being generated, and it is argued in detail in Appendix~\ref{app:spectral} that these contributions are excluded. Therefore, the transmission function is simply
\begin{equation} \label{eq:opt_transf_before_char_eff}
    \mathcal{D}^{\eta}_{\mathrm{opt}}(\varepsilon) = \Theta(i^x(\En))\Theta(i^y(\En)) \Theta\left(\frac{i^x(\En)}{i^y(\En)} - \frac1\lambda\right),
\end{equation}
where the first Heaviside function is included for clarity. From this, we define the characteristic efficiency spectrum
\begin{equation} \label{eq:char_eff}
    \eta^{\mathrm{char}}(\varepsilon) \equiv \frac{i^x(\En)}{i^y(\En)}\Theta(i^x(\En))\Theta(i^y(\En))\ .
\end{equation}
This is the energy-resolved efficiency at full transmission, naturally defined in regimes where both full-transmission spectral currents are positive. One can then identify by Eq.~\eqref{eq:opt_transf_before_char_eff} the inverse Lagrange multiplier as the threshold 
to the characteristic efficiency
\begin{equation}\label{eq:MinCharEff}
    \eta_{\mathrm{min}}^{\mathrm{char}} \equiv 1/\lambda. 
\end{equation}
With this, we express the optimal transmission function in terms of a condition on the characteristic efficiency
\begin{equation}\label{eq:opt_eff_char}
    \mathcal{D}^{\eta}_{\mathrm{opt}}(\En) = \Theta(\eta^{\mathrm{char}}(\varepsilon) - \eta_{\mathrm{min}}^{\mathrm{char}})\ .
\end{equation}
Expressing the result in this way gives an intuitive interpretation: to maximize the efficiency at a given output power, one needs to transmit those particles that contribute \textit{most efficiently} to the total fixed output current. Via the minimum characteristic efficiency, Eq.~\eqref{eq:MinCharEff}, the magnitude of the desired output current decides the lowest acceptable efficiency at each energy. This is shown in Fig.~\ref{fig:char_eff} for the example of cooling using the nonthermal distribution of Fig.~\ref{fig:nonthermal_results}(i). More details of the crossing-point analysis can be found in Appendix~\ref{app:crossings}.

Note that by virtue of $\eta^{\mathrm{char}}(\varepsilon)$ describing the ``spectral'' efficiency, the Lagrange multiplier $\lambda=1/\eta_{\mathrm{min}}^{\mathrm{char}}$ is bounded from below by the inverse of the maximum possible efficiency, which depends on the chosen output and resource currents (the maximal local and total efficiencies coincide). It is unbounded from above as $\lambda \rightarrow \infty \implies \eta_{\mathrm{min}}^{\mathrm{char}} \rightarrow 0$.

\subsection{Lowest noise at fixed output current}\label{sec:gen_noise}
While it is highly relevant that output currents are created at high efficiencies, it is also important for quantum and nanoscale thermoelectrics that fluctuations are small, in other words that the precision of the output current is high. In this section, we hence consider the zero-frequency noise $S^x$ of a generic current $I^x$, see Eq.~\eqref{eq:noise_full}, while keeping the average of this current $I^x$ fixed. The goal is to minimize the noise-to-signal ratio at fixed output current. 

Technically, there is an important difference between the minimization of the noise and the optimization of average currents: the noise has a quadratic contribution in $\mathcal{D}(\varepsilon)$.  
Note however that the shot-noise term $S^x_{\mathrm{shot}}$, which contains all quadratic dependence in $\mathcal{D}(\varepsilon)$, is never negative (it hence always increases the noise) and that the factor $\mathcal{D}(\varepsilon)(1-\mathcal{D}(\varepsilon))$ implies that this term vanishes whenever $\mathcal{D}(\varepsilon) = 0,1$. 
The thermal-like noise in turn is linear in the transmission probability and we can hence apply the same procedure to minimize it as applied in the previous section for the optimization of the efficiency. 
Since the shot noise is suppressed for a boxcar-shaped transmission, and since the thermal noise is linear in $\mathcal{D}(\varepsilon)$, we can anticipate that the optimal transmission for the minimization of noise will be a series of boxcars. Thus, the optimization can be done just on $S^x_{\mathrm{thermal}}$ in a similar fashion as for the efficiencies above.
More rigorous derivations of the fact that it is sufficient to minimize the thermal noise can be found in Appendix~\ref{app:only_thermal} and in Ref.~\cite{Landi2025}. Briefly, the terms quadratic in $\mathcal{D}(\En)$ are concave, such that the minimum points of its derivative are at the boundaries like in the linear case, see Fig.~\ref{fig:linear_illus}(b).

Minimizing the noise-to-signal ratio at fixed current corresponds to minimizing the noise under the constraint of a fixed current. We hence start from the constrained equation
\begin{equation} \label{eq:discrete_noise_constraint}
    S^x_\mathrm{thermal} = \sum_\gamma S^x_{\mathrm{thermal},\gamma}(d_\gamma) +\lambda \big(I^x_\mathrm{fix} - \sum_\gamma I^x_\gamma (d_\gamma)\big).
\end{equation}
The procedure is completely analogous to the one described in Sec.~\ref{sec:gen_eff} with the replacement $y(\varepsilon)(\occdiff) \rightarrow x^2(\varepsilon) (g(\varepsilon) g^-(\varepsilon) + f(\varepsilon) f^-(\varepsilon))$. It follows that the condition for setting $\mathcal{D}(\En) = 1$ is
\begin{eqnarray}
    x^2(\varepsilon) (g(\varepsilon) g^-(\varepsilon) + f(\varepsilon) f^-(\varepsilon))& & \\
    -\lambda x(\varepsilon) (\occdiff)& < & 0 .\nonumber
\end{eqnarray}
Since the first term is strictly positive, the optimal transmission function can be expressed as
\begin{equation}\label{eq:opt_noise_transmission}
    \mathcal{D}^{S}_{\mathrm{opt}}(\varepsilon) = \Theta\left( \lambda-\frac{x(\varepsilon) [g(\varepsilon) g^-(\varepsilon) + f(\varepsilon) f^-(\varepsilon)]}{\occdiff} \right), 
\end{equation}
which is hence the function that minimizes the noise of a generic \textit{fixed} output current. The Lagrange multiplier $\lambda$ is again found from the constraint $I^x\equiv I^x_\mathrm{fix}$. 
In the case where the contacts are thermal and the noise of charge or energy currents is considered, the result of Eq.~\eqref{eq:opt_noise_transmission} reduces to the recent result on precision optimization of Ref.~\cite{Landi2025}.

Analogously to the characteristic efficiency in the previous section, we define the characteristic noise-to-signal ratio
\begin{equation}
    \mathrm{NSR}^{\mathrm{char}}(\En) = \frac{s^x(\En)}{i^x(\En)} \Theta(i^x(\En)) ,
\end{equation}
using the full-transmission spectral current, Eq.~\eqref{eq:spectral_current_1}, and spectral noise, Eq.~\eqref{eq:spectral_noise}. Then, the maximal characteristic noise-to-signal ratio, $\lambda\equiv \mathrm{NSR}^{\mathrm{char}}_\mathrm{max}$, bounds the factor in the Heaviside function of Eq.~\eqref{eq:opt_noise_transmission} from above in order to achieve high precision at given output power,
\begin{equation}
    \mathcal{D}^{S}_{\mathrm{opt}}(\varepsilon) = \Theta( \mathrm{NSR}^{\mathrm{char}}_\mathrm{max}-\mathrm{NSR}^{\mathrm{char}}(\En)).
\end{equation}
The intuitive interpretation is here that in order to maximize the precision one needs to the transmit electrons  which contribute to the output current with the least characteristic noise-to-signal ratio. 

This ratio is bounded from below by 0, but can be arbitrarily large. Since $i^x(\En) >0$ is considered beneficial, only $\mathrm{NSR}^{\mathrm{char}}_\mathrm{max} >0$ gives a positive output current. But $i^x(\En)$ can be arbitrarily close to zero while positive, which always happens near crossing points between the distributions, see Eq.~\eqref{eq:spectral_current_1}. Thus, $\mathrm{NSR}^{\mathrm{char}}_\mathrm{max} \to \infty$ as the $\mathcal{D}^{S}_{\mathrm{opt}}(\varepsilon)$ approaches the transmission function for maximum output, Eq.~\eqref{eq:opt_max_transmission}. Furthermore, $\mathcal{D}^{S}_{\mathrm{opt}}(\varepsilon)$ tends to exclude areas close to where $i^x(\En) \to 0$. 

\subsection{Best precision-efficiency trade-off for fixed output current}\label{secTradeoff}

We have shown in the previous sections how to identify transmission probabilities that result in the best efficiency and how to identify transmission probabilities that result in the lowest noise at a given output current. Our previous arguments show that also for nonthermal distributions, the transmission probabilities that optimize these two goals need to be different, see in addition the discussion in \ref{sec:optimal_fridge}. Indeed, while the efficiency is best when regions of finite transmission select the crossing points between distribution functions~\cite{Hicks1993May,Mahan1996Jul}, the precision is best when the distribution functions differ maximally in the transport window selected by the transmission probability~\cite{Palmqvist2024}. As a next step, we therefore aim to optimize the trade-off between the two. Such a trade-off can for example be formulated in terms of a thermodynamic uncertainty relation (TUR), see Eq.~\eqref{eq:TUR_def}.

We proceed as in the previous sections and optimize general trade-off coefficients as defined in Eq.~\eqref{eq:TUR_def}, again incorporating the normalization factor $c_A$ in the noise $S^x$.  
When taking the output current $I^x$ to be fixed, the optimization of the expression in Eq.~\eqref{eq:TUR_def} reduces to an optimization of the product $S^x I^y$ with respect to the transmission function.
Since the quantity to be optimized is a product of two integrals, the results from the previous sections can not simply be transferred. Instead, we start by applying a variation of $\mathcal{D}(\En)$ on the entire product with the constraint $I^x\equiv I^x_\mathrm{fix}$ included via a Lagrange multiplier,
\begin{eqnarray}
\delta(S^xI^y) - \lambda\delta I^x
& = &     I^y\delta S^x  + S^x\delta I^y - \lambda\delta I^x .
 \label{eq:prod_pre_cond}
\end{eqnarray}
Note that it is sufficient to include the constraint $I^x$ only as linear term in the Langrage function~\eqref{eq:prod_pre_cond} as the goal is to optimize the numerator in the TUR-inspired coefficient~\eqref{eq:TUR_def}.
Similar to the discrete condition in Eq.~\eqref{eq:discrete_cond}, the discretized version of the product variation is
\begin{equation}\label{eq:prod_discrete}
    \left.\frac{\partial S^xI^y}{\partial d_\gamma}\right|_{I^x_\mathrm{fix}} = S^x\frac{\partial I^y}{\partial d_\gamma} +I^y \frac{\partial S^x}{\partial d_\gamma} - \lambda \frac{\partial I^x}{\partial d_\gamma},
\end{equation}
where the differentiation with respect to $d_\gamma$ picks out one energy slice, while the noise $S^x$ and the current $I^y$, which occur as prefactors of the derivatives in Eq.~\eqref{eq:prod_pre_cond}, are sums over all slices. Then the procedure follows as in Sec.~\ref{sec:gen_eff}, with $d_\gamma$ being set to either 0 or 1 depending on the sign of the derivatives, and with $S^x$ replaced by $S^x_\mathrm{thermal}$ as in Sec.~\ref{sec:gen_noise}. The key difference is that the constants $S^x$ and $I^y$ are not arbitrary but must be solved \textit{self-consistently}, similarly to how $\lambda$ must be solved for a specific fixed output current $I^x$.

Going to the continuous limit, the variations in Eq.~\eqref{eq:prod_pre_cond} become energy-resolved, as they determine how $\mathcal{D}(\En)$ changes, while $I^y$ and $S^x$ are kept in their integral form\footnote{In essence the method is the same as for optimizing the efficiency with the replacement $y(\varepsilon)(\occdiff) \rightarrow -I^y x^2(\varepsilon)(g(\varepsilon) g^-(\varepsilon) + f(\varepsilon) f^-(\varepsilon)) - S^xy(\varepsilon)(\occdiff)$.}. The optimal transmission function is then
\begin{eqnarray}    \label{eq:opt_transmission_product}
    \mathcal{D}^A_\mathrm{opt}(\varepsilon) = \Theta\Big(-I^y s^x(\En)
    - S^xi^y(\En) + \lambda I^x(\En)\Big)\ .
\end{eqnarray}

Crucially, in order for this result to hold, each of the three functions $S^x$, $I^y$, and $I^x$ must be solved for, which is accomplished by inserting $\mathcal{D}^A_\mathrm{opt}(\En)$ into their integral equations, see Eqs.~\eqref{eq:generic_current} and~\eqref{eq:noise_full}, and solving for all three simultaneously.

Following the same reasoning as in Sec.~\ref{sec:gen_eff} and Appendix \ref{app:spectral}, the spectral currents should be positive and therefore $\lambda >0 $. Similar to the efficiency and noise-to-signal ratio, we can rewrite the condition as
\begin{eqnarray}\label{eq:alt_opt_transmission_product}
    \mathcal{D}^A_\mathrm{opt}(\varepsilon) &=& \Theta\left(\lambda - \frac{I^ys^x(\En) + S^xi^y(\En) }{i^x(\En)}\right) \nonumber \\ &\times&\Theta(i^x(\En))\Theta(i^y(\En)). 
\end{eqnarray}

In contrast to the separate efficiency and noise-to-signal optimizations, there is not a clear parallel characteristic quantity to Eq.~\eqref{eq:TUR_def} which can be used to interpret the transmission function, since the numerator in Eq.~\eqref{eq:alt_opt_transmission_product} contains both spectral and integrated functions. However, this sum serves as a proxy of the product $s^x(\En)i^y(\En)$.

Logically, the energy windows transmitting electrons should be the ones that contribute the least to $A^{x,y}$ and $\lambda$ sets the threshold for their maximal contribution. 

An interesting consequence of this type of optimization of a product is that one can straightforwardly introduce an additional weight $\alpha \in [0,1]$,
\begin{equation} \label{eqAlphaResult}
    (2\alpha\delta S^x I^y + 2(1-\alpha)S^x\delta I^y - \lambda\delta I^x) < 0 .
\end{equation}
This equation directly yields the transmission function
\begin{align}
\mathcal{D}_\mathrm{opt}^\alpha(\varepsilon) = \Theta(-2\alpha I^y s^x(\varepsilon)
    - 2(1-\alpha )S^xi^y(\varepsilon) + \lambda i^x(\varepsilon)) .          
\end{align}

Such a hypervariable $\alpha$ could tune to which extent the precision should be weighted compared to the efficiency in a chosen trade-off relation. For $\alpha = 0$ the result in Sec.~\ref{sec:gen_eff} is regained by casting $\lambda/(2S^x)\rightarrow\lambda$, keeping in mind that $S^x$ can be considered as a constant. If instead $\alpha = 1$, then the condition is equivalent to the result in Sec.~\ref{sec:gen_noise}, with $\lambda/(2I^y) \rightarrow\lambda$. Setting $\alpha = 1/2$ just gives back the product relation for $A^{x,y}$.
\subsection{Optimization with respect to a second constant}\label{sec:opt_two}
We have until here optimized the performance of the energy-conversion process by tuning the transmission function of the conductor. In many cases, it might be relevant to optimize with respect to another parameter in addition to the transmission function. For example, for \textit{thermal} reservoirs it is useful to consider the potential difference as a variable, as it can typically be externally modified. In earlier results~\cite{Whitney2014Apr,Whitney2015} for the optimization of efficiency at fixed power, the variation in terms of the potential difference was directly used when determining that the optimal transmission function was a boxcar. In the following we present a method that mirrors very closely the procedure in Ref.~\cite{Whitney2015}, but considering any \textit{arbitrary} second variable. 

\begin{figure}[tb]
    \centering
    \includegraphics[width=\linewidth]{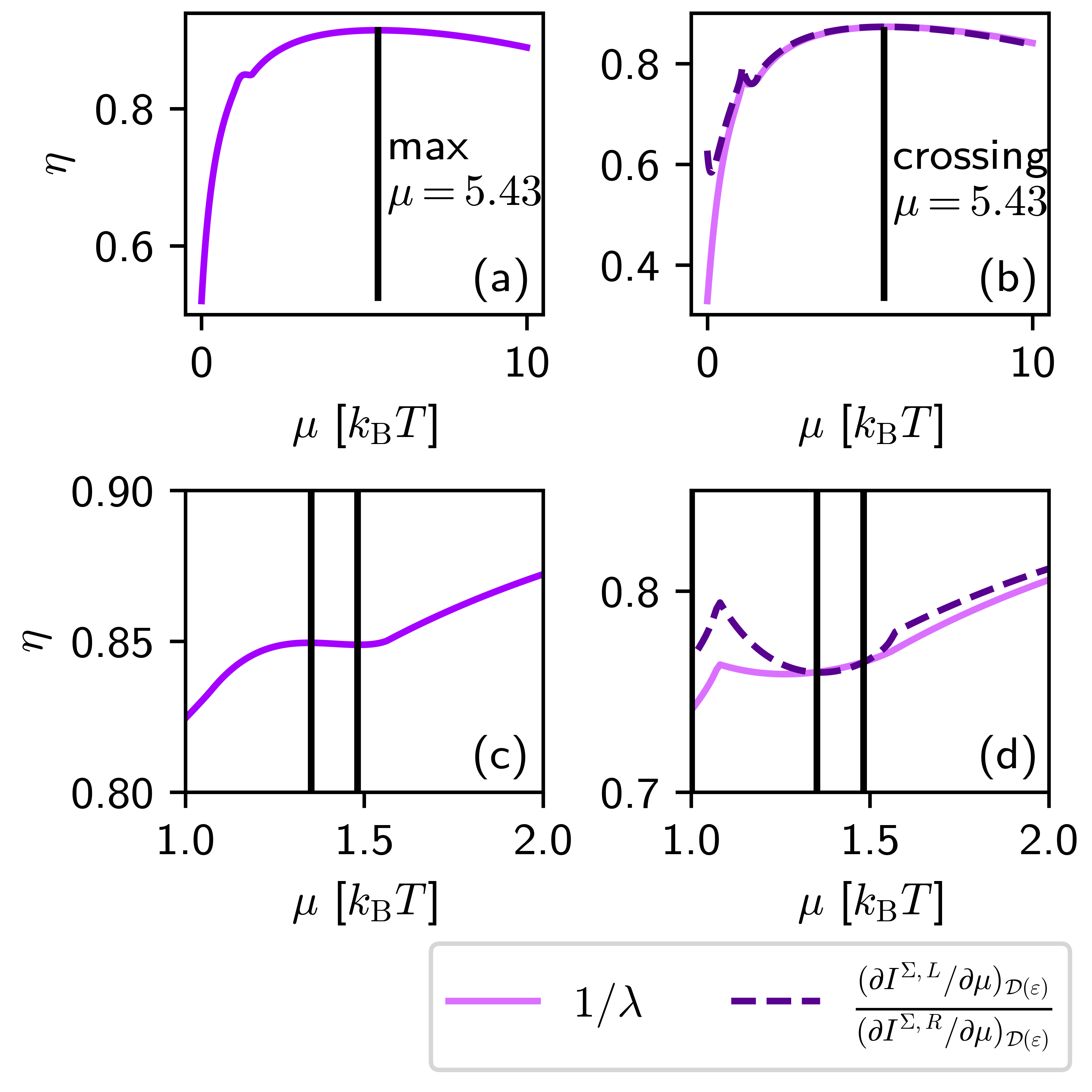}
    \caption{ Optimized efficiencies for the fixed left entropy current $I^{\Sigma, \mathrm{L}} = 0.15\: k_\mathrm{B}^2 T/h$ by varying the right entropy current $I^{\Sigma, \mathrm{R}}$ over a range of $\mu \in [0,10] \ k_\mathrm{B}T$, between the cold thermal distribution in L and a nonthermal distribution with a dip-peak structure. (a) shows the calculated efficiencies with optimal transmission functions for each $\mu$, with the maximal efficiency found in $\mu = 5.43 \ k_\mathrm{B}T$. In (b), the light purple curve shows $1/\lambda$ associated with the optimal transmission functions at each $\mu$ and the dark purple, dashed curve is the calculated fraction of partial derivatives as found in Eq.~\eqref{eq:avg_second}. (c) and (d) are zooms of the upper panels, picking our the region $\mu \in [1,2]\ \kB T$ where additional extrema occur (indicated by black veritcal lines).}
    \label{fig:mu_example}
\end{figure}

We hence choose a second variable $\xi$, which is independent of the energy $\varepsilon$. We consider as our goal the maximization of the efficiency with a fixed output current $I^x_\mathrm{fix}$, like in Sec.~\ref{sec:gen_eff}, but now by varying both $\mathcal{D}(\En)$ \textit{and} $\xi$. The approach is to solve for the unknown $\lambda$ by introducing a second constraint equation. Since $\xi$ is not energy-dependent, the variations of the currents with respect to it are direct partial derivatives, and as such they are not energy-resolved. We find
\begin{subequations} \label{eq:secondary_constraint}
    \begin{eqnarray}
        \delta_\mathcal{D}I^y|_\xi - \lambda\delta_DI^x|_\xi &<& 0, \label{eq:secondary_constraint_a}\\
        \left.\frac{\partial I^y}{\partial \xi}\right|_\mathcal{D} - \left.\lambda\frac{\partial I^x}{\partial \xi}\right|_\mathcal{D} &=& 0. \label{eq:secondary_constraint_b}
    \end{eqnarray}
\end{subequations}
The second line can be solved for $\lambda$ which is then inserted into the first line, giving
\begin{equation} \label{eq:avg_second}
    \left.\delta_\mathcal{D} I^y\right|_\xi - \frac{\partial I^y / \partial \xi |_\mathcal{D}}{\partial I^x / \partial \xi|_\mathcal{D}} \left.\delta_\mathcal{D} I^x\right|_\xi < 0 .
\end{equation}
For terminals with thermal distributions and with $\xi \equiv \Delta \mu$, this condition coincides with the result of Ref.~\cite{Whitney2015}. Effectively, in the situation where the efficiency is optimized by tuning $\mu$ in addition to $\mathcal{D}(\En)$, the unknown constant $\lambda$ in previous sections is replaced by the derivatives of the currents with respect to the second variable, \textit{evaluated at fixed transmission function} defined by the full transmission condition. 
In other words, the optimal $\xi$ for a certain value of $I^x$ is found at the point where the Lagrange multiplier $\lambda$ for a given $\xi$ and the fraction of derivatives $\frac{\partial I^y / \partial \xi |_\mathcal{D}}{\partial I^x / \partial \xi|_\mathcal{D}}$ coincide. We visualize this in Fig.~\ref{fig:mu_example} for the example of finding the optimal efficiency for a certain fixed current (here $I^x\to I^{\Sigma,\mathrm{L}}$) over a range of $\mu$, which here represents a shift in the electrochemical potential of the cold reservoir compared to the zero of energy with respect to which the nonthermal distribution in terminal R is defined. Panel (a) shows the result of the efficiency optimized by ``brute-force" with respect to $\mu$, by finding the optimal transmission function for the same $I^{\Sigma,\mathrm{L}}_\mathrm{fix}$ over a range of $\mu$. This result perfectly coincides with the crossing of $\lambda(\mu)$ and the calculated fraction of derivatives, as sketched in Fig.~\ref{fig:mu_example}(b). 

However, note that the constraint in Eq.~\eqref{eq:secondary_constraint_b} is an equality, as opposed to an inequality like Eq.~\eqref{eq:secondary_constraint_a}. This means that there can be several solutions that fulfill the criterion $I^x_\mathrm{fix}$, if $I^y$ has several extrema with respect to $\xi$ and optimized transmission functions. This is illustrated in panels (c) and (d) in Fig.~\ref{fig:mu_example}. In panel (c), there is a local maximum and minimum, in which Eq.~\eqref{eq:secondary_constraint_b} is fulfilled. This is clear in panel (d), where $\lambda$ is shown to coincide with the fraction of derivatives at both points. Thus, the more favorable solution should be selected, if co-existing solutions exist.

As a consequence of this additional optimization step, the efficiency is higher than the one found when optimizing with  a fixed $\lambda$ (and hence a fixed $\mu$). 

This procedure can straightforwardly be extended to the general optimization of a product of functions, presented in Sec.~\ref{secTradeoff}. Here, one finds for the Lagrange multiplier from Eq.~\eqref{eqAlphaResult} for a general $\alpha$,
\begin{equation}
    \lambda = \frac{2\alpha I^y\partial S^x / \partial \xi|_\mathcal{D} + 2(1-\alpha)S^x \partial I^y/\partial \xi|_\mathcal{D}}{\partial I^x / \partial \xi|_\mathcal{D}}\ .
\end{equation}
This is hence an analogous problem to the one we solved in Sec.~\ref{secTradeoff}, but now $\lambda$ is determined by the derivatives of the currents and fluctuations with respect to the second parameter $\xi$, at a fixed transmission function. 
For the specific values of $\alpha =0, 1/2, 1$, corresponding to an optimization of efficiency, TUR-like trade-off, and low noise-to-signal ratio at fixed output current, one finds
\begin{subequations}
    \begin{eqnarray} 
        \alpha &=& 0 \rightarrow \lambda = \frac{\partial I^y / \partial \xi|_\mathcal{D}}{\partial I^x / \partial \xi|_\mathcal{D}} \label{eq:second_eff}, \\ 
        \alpha &=& 1/2 \rightarrow \lambda = \frac{\partial S^x / \partial \xi|_\mathcal{D}}{\partial I^x / \partial \xi|_\mathcal{D}}, \\
         \alpha &=& 1 \rightarrow \lambda  = \frac{\partial (S^xI^y) / \partial \xi|_\mathcal{D}}{\partial I^x / \partial \xi|_\mathcal{D}}.
    \end{eqnarray}
\end{subequations}
\section{Optimized cooling exploiting a nonthermal resource}\label{sec:optimal_fridge}
\begin{figure*}[tb]
    \subfloat{\label{fig:dippeak_results}
        \includegraphics[width = 0.45\linewidth]{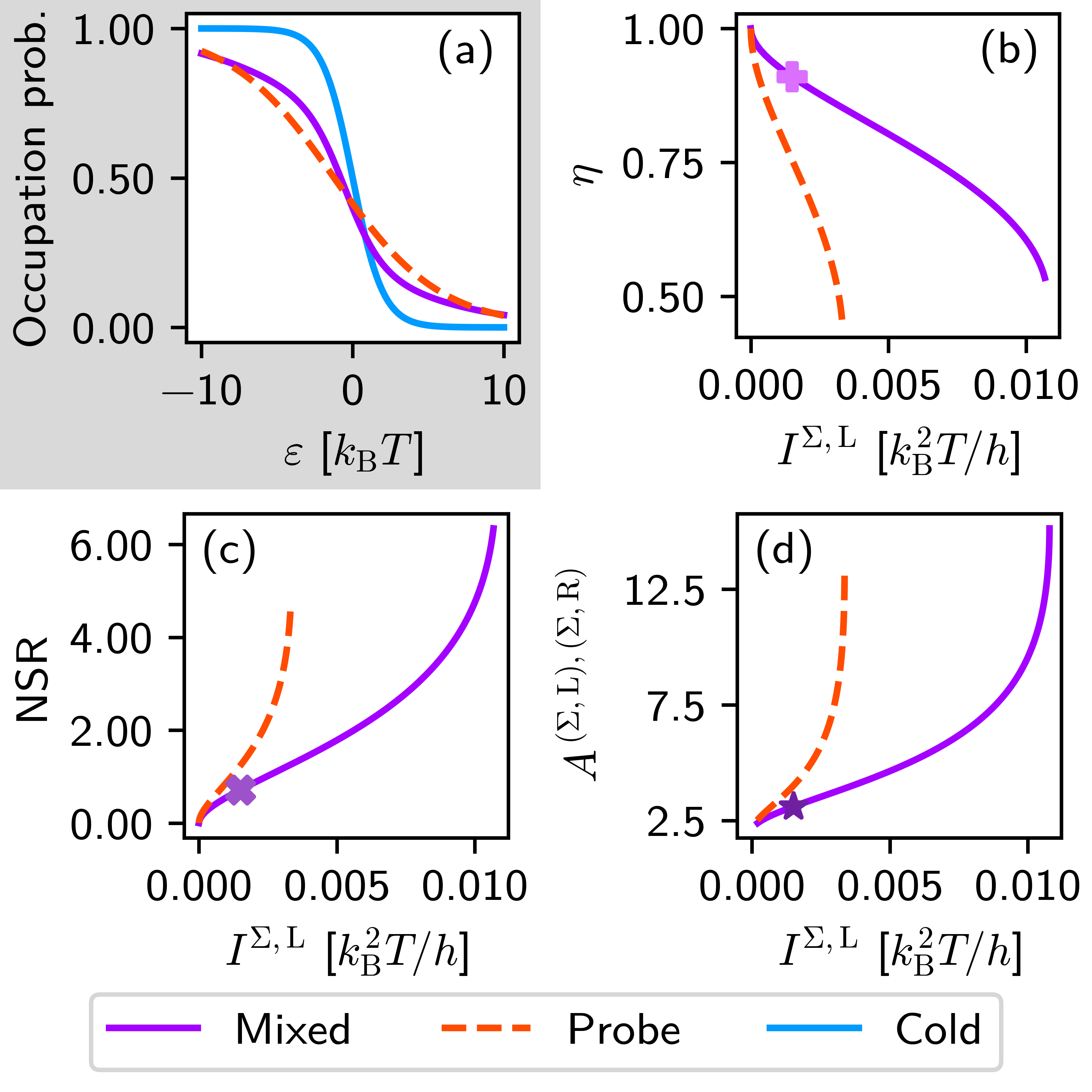}}
    \hfill
    \subfloat {\label{fig:mixed_results}
        \includegraphics[width = 0.45\linewidth]{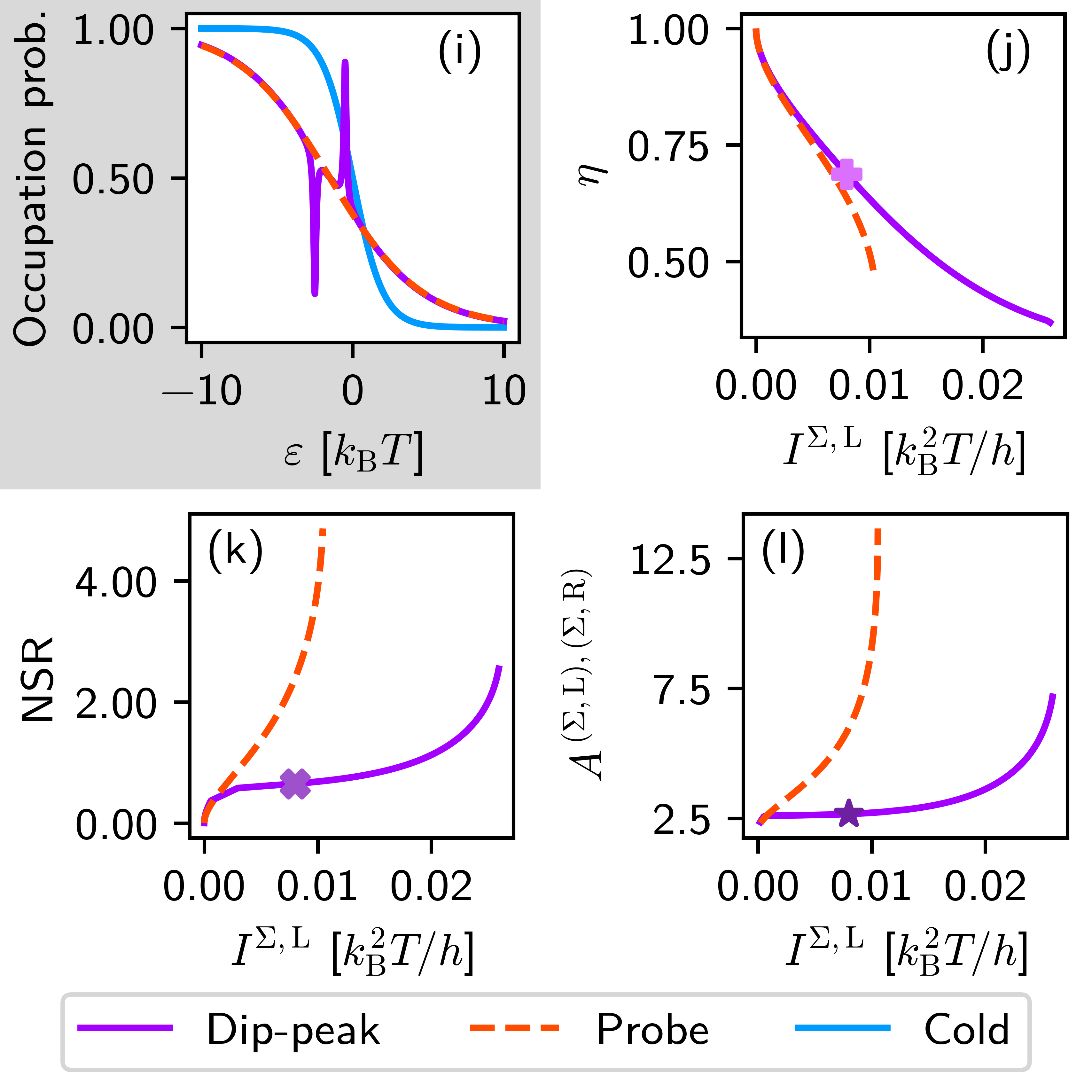}}
    \hfill
    \subfloat{\includegraphics[width = 0.45\linewidth]{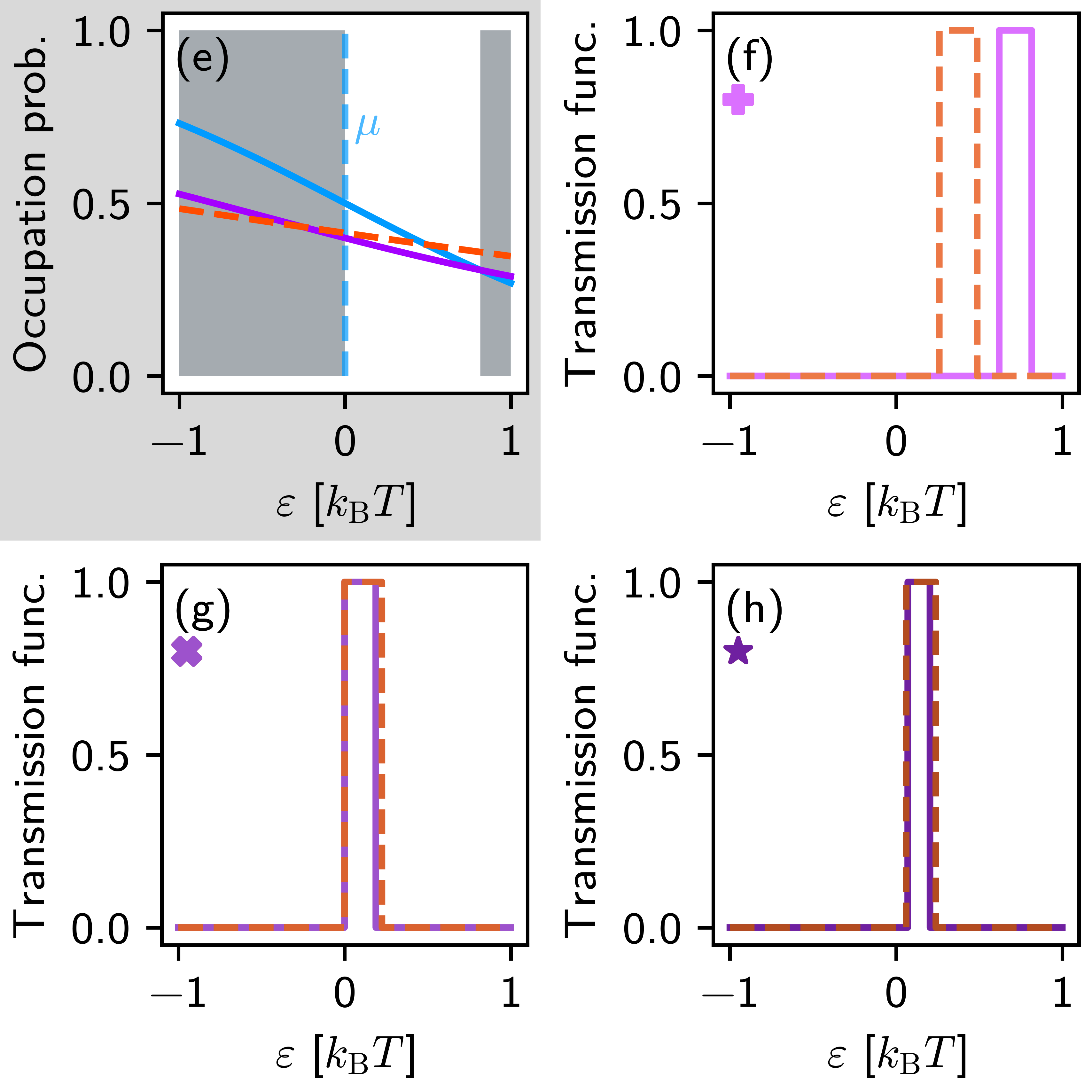}}
    \hfill
    \subfloat{\includegraphics[width = 0.45\linewidth]{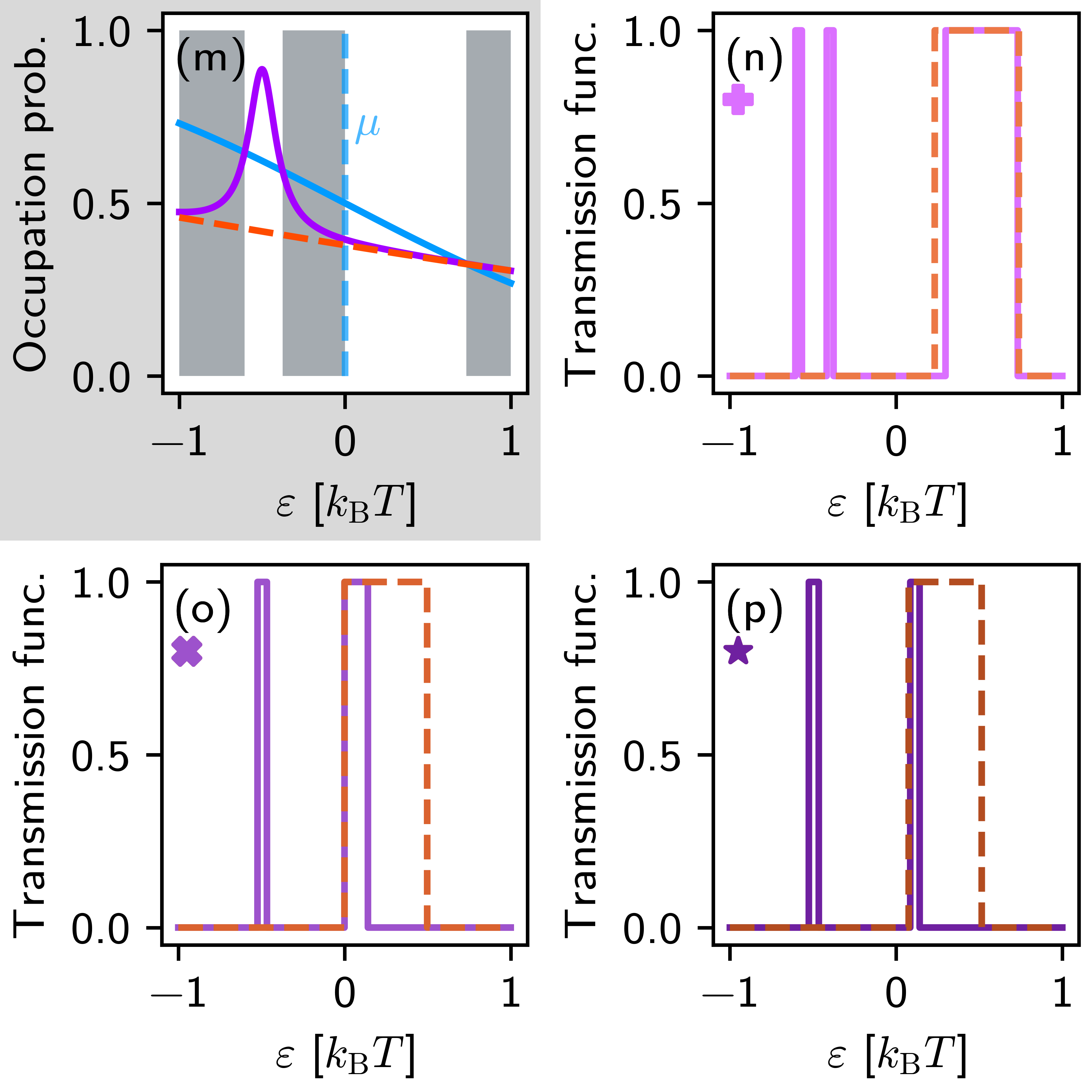}}
    
    \caption{Optimized performance for concrete nonthermal distributions. (a) Mixed distribution, Eq.~\eqref{eq:mixed_distribution} and its thermal equivalent, both used to obtain (b) optimal efficiency (c) optimally low noise-to-signal ratio and (d) precision-efficiency trade-off. (e) Zoom-in of (a) where non-cooling regions of the nonthermal distribution are shaded in gray. (f) to \correct{(h)} show optimal transmission functions at the fixed cooling power indicated by crosses and stars in (b) to (d). (i) Dip-peak distribution, Eq.~\eqref{eq:dip_peak_distribution} and its thermal equivalent, both used to obtain (j) optimal efficiency (k) optimally low noise-to-signal ratio and (l) precision-efficiency trade-off. (m) Zoom-in of (i) where non-cooling regions of the nonthermal distribution are shaded in gray. (n) to (p) show optimal transmission functions at the fixed cooling power indicated by crosses and stars in (j) to (l). }
    \label{fig:nonthermal_results}
\end{figure*}
We have until here introduced an optimization procedure for general output currents, resources, efficiencies and precision for a generic steady-state two-terminal coherent conductor described by scattering theory. 
In this section, we apply the developed procedure to a relevant thermodynamic process, namely cooling. The task to be completed is hence to achieve a positive steady-state cooling power $ I^{Q,\mathrm{L}}$ such that $I^x \to I^{Q,\mathrm{L}}/T $, while using the nonthermal distribution in the right contact as resource and the transmission properties of the quantum conductor as energy filter. The resource ``used up" can be quantified by the entropy production in the right (nonthermal) contact $I^y\to - I^{\Sigma,\mathrm{R}}$. The sketch of Fig.~\ref{fig:Setup} illustrates the working principle. 

In this section, we  consider two different nonthermal distributions as experimentally relevant examples: (A)~a mixture of different thermal distributions as it results from the interplay of competing environments and (B)~a distribution with peaks and dips as it could result from a fixed-frequency irradiation on a thermal background. The latter one has been used in the examples of the previous section.

In order to compare the opportunities arising from a nonthermal distribution as a resource to those from a thermal resource, we use the concept of B\"uttiker probes~\cite{Buttiker1985Aug,Buttiker1986Mar,Buttiker1988Nov,deJong1996Aug,Forster2007Jan} to assign an \textit{effective} temperature $T_\mathrm{eff}$ and an \textit{effective} electrochemical potential $\mu_\mathrm{eff}$ to the probe distribution. A B\"uttiker probe can be imagined as a floating contact attached to the terminal with the nonthermal distribution, where the probe adjusts its temperature and electrochemical potential under the condition that on average no particles and no energy is transferred between the nonthermal terminal and the probe \footnote{We choose a standard measure to define the effective temperature and chemical potential to the average charge and energy current while other measures might consider, e.g., the noise in the particle current. See~\cite{Giazotto2006Mar} for an overview on thermometry on the nanoscale.}. 

In the following, we analyze the performance of cooling with the two example nonthermal distributions. We determine the series of boxcar transmissions that is required to optimize the performance as quantified by cooling power, efficiency, precision and a trade-off relation. We finally show in Sec.~\ref{sec:lorentzian_transmission} how the performance is modified, when the ideal boxcar transmissions are replaced by Lorentzian transmissions, which is one commonly occurring transmission function stemming from resonances in experimentally realized conductors.
For details of the numerical procedures employed, see Appendix~\ref{app:numerical} and Ref.~\cite{zenodo}. 

\subsection{Nonthermal distribution from competing environments}\label{sec:mixed}

We start by presenting an example of a nonthermal distribution consisting of a mixture of two thermal distributions, henceforth referred to as a ``mixed distribution''. Such a distribution can result from the competition of different environments (such as sun light and phonon background in solar cells~\cite{Tesser2023} or from two mixed electronic distributions at different potentials in a quantum Hall setup~\cite{leSueur2010Jul,Altimiras2010Nov}, to name two examples). We construct the nonthermal distribution from 
\begin{equation}\label{eq:mixed_distribution}
g(\En) = a f_{\mathrm{env},1}(\En) + b f_{\mathrm{env},2}(\En)\ .
\end{equation}
with $a+b=1$ and with the thermal distributions of the competing environments $f_{\mathrm{env},i}(\En)$ at temperatures $T_{\mathrm{env},i}$ and electrochemical potentials $\mu_{\mathrm{env},i}$. 
The resulting distribution, $g(\En)$, is shown as purple line in Fig.~\ref{fig:nonthermal_results}(a) for $a=b=1/2$ and for $\mu_{\mathrm{env},1}=-2k_\mathrm{B}T$ and $T_{\mathrm{env},1}=5T$, $\mu_{\mathrm{env},2}=-0.5k_\mathrm{B}T$ and $T_{\mathrm{env},2}=1.2T$ , where the nonthermal nature is seen in the ``step'' which is not present in thermal distributions. The hot thermal distribution that is found from connecting the nonthermal contact to a B\"uttiker probe has 
$T_\mathrm{eff}= 3.5035 \:T$ and $\mu_\mathrm{eff}=-1.2214\:k_\mathrm{B}T$ (rounded to the fourth decimal),
and it is shown as the red line in the same panel. The cold distribution in the left contact has temperature $T$ and electrochemical potential $\mu=0$. 

We show the results for optimal efficiency, precision and noise-efficiency trade-off, see Eqs.~\eqref{eq:opt_eff_transmission},~\eqref{eq:opt_noise_transmission}, and~\eqref{eq:opt_transmission_product}, for the nonthermal distribution (purple) and the equivalent thermal probe distribution (red) coupled to the cold distribution
in Fig.~\ref{fig:nonthermal_results}(b)-(d). Note that the transmission functions of the coherent conductor are optimized independently  for the nonthermal distribution and for its thermal equivalent, which means that applying the optimal transmission function for either case to the other would invariably give a worse performance.

The end points of the different curves correspond to the highest achievable cooling power $\propto I^{\Sigma,\mathrm{L}}$. It is important to note that this  highest achievable cooling power is significantly larger in the nonthermal case compared to the thermal equivalent.
The reason for this can be seen in the zoom of the distribution function in panel (e): in the energy interval above $\mu=0$, where the right distribution is below the cold, "hot" electrons can be extracted from the cold, left distribution, resulting in a positive cooling power. This interval is larger for the nonthermal distribution compared to its thermal equivalent, leading to a larger possible cooling interval.

Remarkably, the nonthermal distribution performs better for every quantifier and at any value of the fixed output current  $I^{\Sigma, L}$, proving that nonthermality is worth considering when designing nanoscale devices. 

The reason for this improved performance can be understood from the transmission functions that lead to the optimal values of the performance quantifiers. For a fixed cooling power $I^{\Sigma,\mathrm{L}}=0.0015 \:k_\mathrm{B}^2T/h$ indicated by crosses and stars in panels (b) to (d), these transmission functions are shown in panels (f) to (h), 
for each performance quantifier and for both the nonthermal distribution and the equivalent thermal probe distribution. 
 The transmission window leading to optimal efficiency for a nonthermal distribution (shown in purple in (f)) lies in a significantly higher energy interval than for the thermal equivalent.  As a result, the same cooling power can be achieved, even though the window of transmission is narrower and even though the occupation probabilities and their difference are lower. The narrower window of transmission together with the reduced difference between the occupations reasonably leads to less entropy production in the resource and hence a higher efficiency.
 
 Interestingly, the transmission found for the optimal efficiency for the nonthermal distribution is located in an energy interval in which the thermal distribution can not achieve cooling at all. A device designed for achieving optimal efficiency in the nonthermal case, would hence have zero efficiency in the thermal case.

Also for the precision and the trade-off, the optimal transmission windows are narrower for the nonthermal distribution compared to the thermal one. 
The reason is that the narrower the transmission window, the smaller the contribution from the spectral noise and from the spectral resource current (which only have positive values in the intervals of interest) tend to be.
This further indicates that it is beneficial to have a narrow transmission window while keeping the fixed current constant. 
 
\subsection{Nonthermal distribution from irradiation}\label{sec:irradiation}

We furthermore consider an example distribution consisting of an underlying thermal distribution with an added Lorentzian peak and dip, see Fig.~\ref{fig:nonthermal_results}(i). Such a dip-peak shape mimics excitations stemming from fixed-frequency light irradiation, promoting electron excitations to higher energies while leaving behind hole-like excitations in the Fermi sea, see for example Refs.~\cite{Song2015Apr,Kim2011Aug}.
This distribution has the interesting feature of \textit{nonmonotonicity} and it will be referred to as the ``dip-peak distribution'' henceforth.  Concretely, it is defined by
\begin{equation}
    g(\En) = f_\mathrm{env}(\En) + r\frac{\pi}{2}\sum_{i=\mathrm{e,h}} \sigma_i\frac{\gamma_i^2}{(\En - \En_i)^2 + \gamma_i^2}     \label{eq:dip_peak_distribution}  
\end{equation}
which creates a dip-peak structure around $\mu_\mathrm{env}$ where $\sigma_i=\pm1$ accounts for the sign of the modifications for electron-and hole-like excitations. The peak and dip are situated around $\mu_\mathrm{env}$ with distance $\En_i$ and with the Lorentzian height $r\pi/2$ with $r = 0.3$ and half-width-at-half-maximum $\gamma_i$. The resulting distribution with $\mu_\mathrm{env}=-1.5 \: k_\mathrm{B} T$ and $T_\mathrm{env}=3 T$, as well as $\gamma_\mathrm{d}=\gamma_\mathrm{p}=0.1$ and $\En_\mathrm{d} - \mu_\mathrm{env}=-\En_\mathrm{p}+\mu_\mathrm{env}=1$  is shown in purple in  Fig.~\ref{fig:nonthermal_results}(i). The dip and peak are symmetrical around $\mu_\mathrm{env}$. Due to this sharp symmetrical feature, the B\"uttiker probe ``sees" an effective distribution which is very close to the underlying thermal distribution, as evidenced by the calculated quantities
$\mu_\mathrm{eff}=-1.5003\:k_\mathrm{B} T$ and $T_\mathrm{eff}= 3.0312\: T$, rounded to the fourth decimal. The distribution is shown as a red line in the same panel.

\subsubsection{Optimization by tuning the transmission function}

We start by showing the results for the optimal efficiency, noise-to-signal ratio, and trade-off parameter between efficiency and precision at fixed cooling power in Fig.~\ref{fig:nonthermal_results}(j)-(l) obtained by optimizing the transmission function $\mathcal{D}(\varepsilon)$ in analogy to the study based on the mixed distribution in  Sec.~\ref{sec:mixed}.
Similar to the results achieved with the mixed distribution, the dip-peak distribution consistently performs better than its thermal equivalent for each of the quantifiers. In particular the \correct{improvements} concerning the precision and trade-off in panels (k) and (l) \correct{stick} out.

This can again be understood by analyzing the transmission functions that optimize the performance as shown in panels (n) to (p), for the fixed current $I^{\Sigma,\mathrm{L}}=0.0008\:k_\mathrm{B}^2T/h$. A crucial difference compared to the results obtained with the mixed distribution results from the nonmonotonicity of the dip-peak distribution: the optimal transmission functions \correct{now} consist of series of boxcar functions---instead of only a single one.  

One clearly observes that the series of boxcars optimizing the efficiency are located around the crossing points between $f(\varepsilon)$ and $g(\varepsilon)$. To understand this, consider the characteristic efficiency in Fig.~\ref{fig:char_eff}, which is highest near the crossing points. At these points, the left and right entropy current coefficients are equal, such that $i^x(\En)/i^y(\En) \to1$.
In contrast, for optimal precision the boxcar transmission functions are centered around intervals where $f(\varepsilon)$ and $g(\varepsilon)$ differ significantly. This can be understood through the characteristic noise-to-signal ratio, which diverges where $i^x(\En) \to 0$. Thus, the precision optimization seeks energy areas far from crossing points. 

The additional crossing points showcase that nonthermal distributions can give additional freedom to the optimization. The optimal transmission functions for the thermal distribution are practically defined by fewer variables: the optimal function for efficiency always starts at the crossing point while for precision it always starts the at the electrochemical potential, the farthest possible \correct{point} from the crossing \correct{point}. Thus, only the other end of the boxcar is determined by the fixed output current. However, for the nonthermal distribution, there are two additional crossing points. The optimal transmission function in panels~(n) and~(m) picks out areas near these, while keeping the transmission window narrow in the cooling window \correct{shared by} the nonthermal distribution and its thermal equivalent. The trade-off optimization also exploit the new cooling windows around the nonmonotonicity. 

Thus, we can conclude that nonmonotonicities can be beneficial since they introduce more options for placing the transmission boxcars. 

\subsubsection{Optimization with $\mu$ as second variable}
We demonstrate the consequences of optimizing cooling by modulating both the transmission function $\mathcal{D}(\varepsilon)$ and the electrochemical potential $\mu$, solving for the optimal point as outlined in Sec.~\ref{sec:opt_two}. We do this at the example of optimal efficiency at a given cooling power exploiting a dip-peak nonthermal distribution. Essentially, one finds $\lambda$ such that Eq.~\eqref{eq:second_eff} is fulfilled for a transmission function determined by Eq.~\eqref{eq:opt_eff_transmission} where we choose $\mu\in[0,10]k_\mathrm{B}T$ to focus on one of the peaks in the distribution.\footnote{Choosing negative values for $\mu$ would result in a similar behavior deriving from the \textit{dip} in the distribution. Higher values of $|\mu|$ are not of interest since the nonthermal properties of the dip-peak distribution would not be visible any longer.}  
\begin{figure}[bt]
    \centering
    \includegraphics[width=\linewidth]{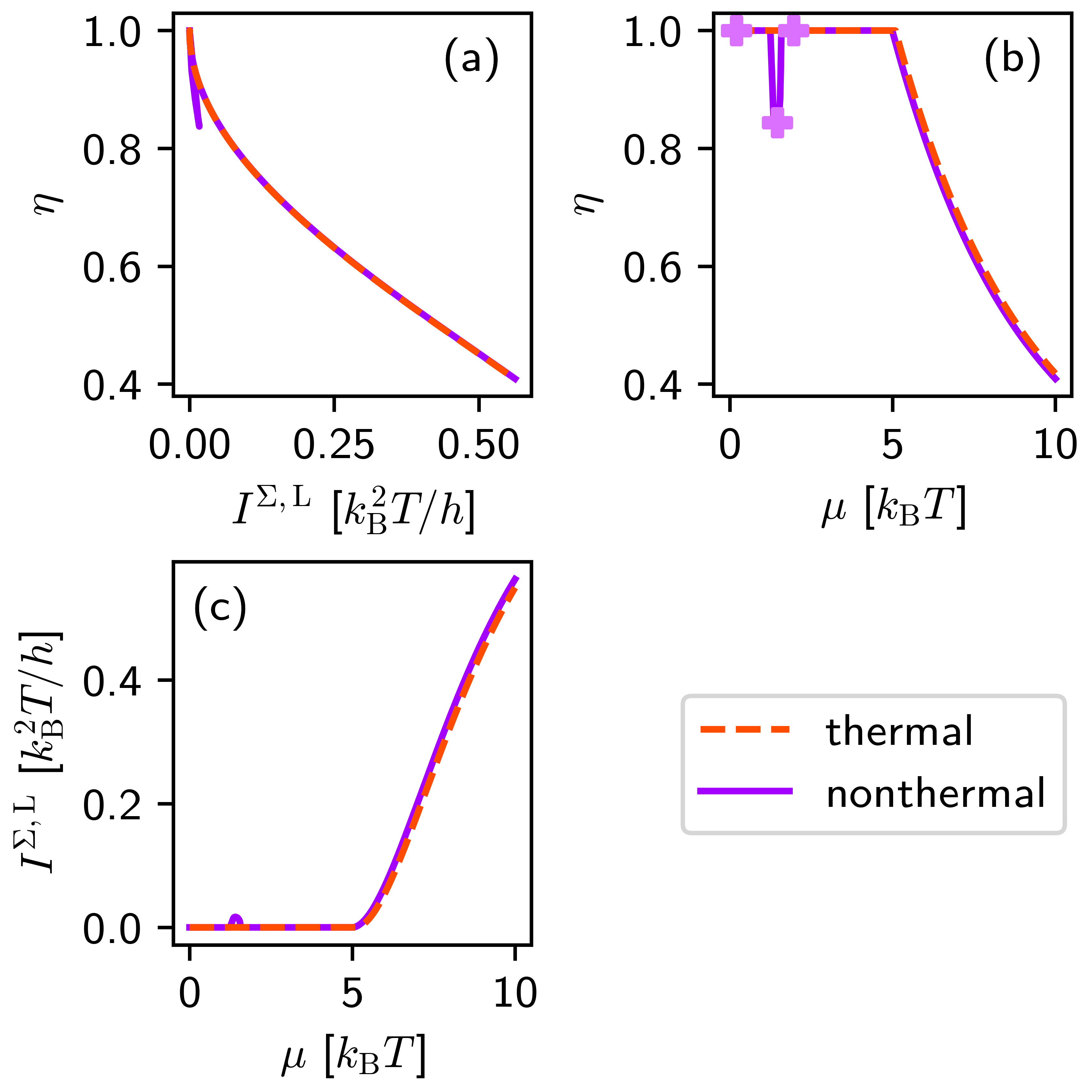}
    \caption{(a) Optimal efficiency, obtained from the nonthermal dip-peak distribution and from its thermal equivalent by optimizing the transmission function $\mathcal{D}(\varepsilon)$ as well as the electrochemical potential of the cold thermal distribution $f(\varepsilon)$, as function of the fixed cooling power $I^{\Sigma,\mathrm{L}}$. 
    (b) Optimal efficiency as function of $\mu$, which optimizes the efficiency at each fixed cooling power. \correct{The three pluses indicate the $\mu$s that are used in the three examples in Fig.~\ref{fig:secondary_transition}.}(c) Relation between fixed cooling power $I^{\Sigma,\mathrm{L}}$ and the electrochemical potential $\mu$ leading to optimal efficiency at $I^{\Sigma,\mathrm{L}}$.}
    \label{fig:secondary_results}
\end{figure}
\begin{figure}[bt]
    \centering
    \includegraphics[width=\linewidth]{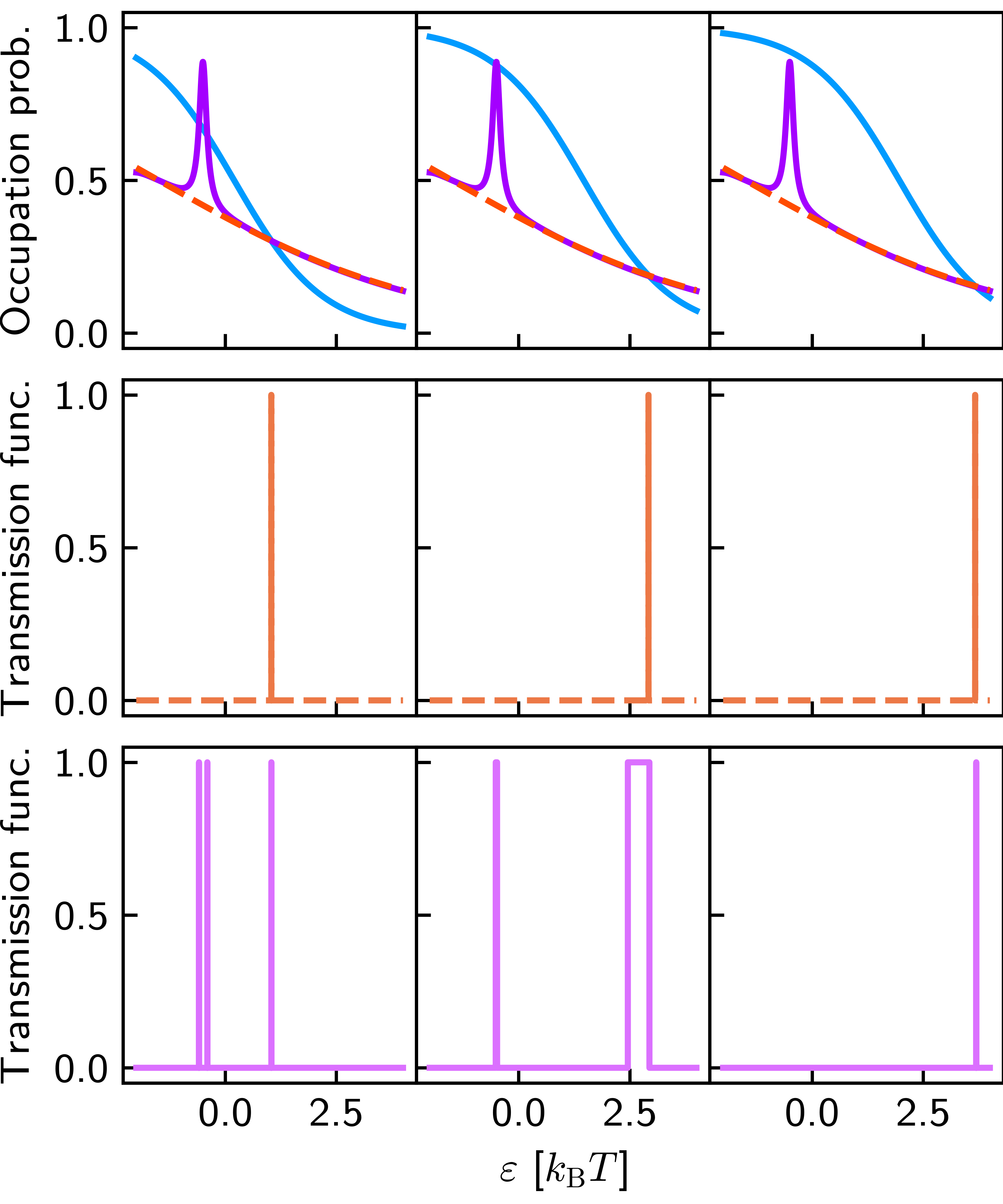}
    \caption{Solutions to optimization with second variable as the cold distribution (blue line) passes the peak in the dip-peak distribution (purple line). The first row shows the distributions with increasing $\mu$ from left to right. The second row shows the thermal optimization, corresponding to the red line in the first row. These transmission functions are all very narrow boxcars near the crossing point between the cold and thermal distributions. The third row shows the optimal transmission functions for the nonthermal distribution. The first and third panel show narrow transmission windows near the crossing points. The second panel showcases one of the co-existing solutions, with $I^{\Sigma, \mathrm{L}} = 0.015\:\kB T$. All other resulting currents or of order $10^{-9}\:\kB T$ or less.}
    \label{fig:secondary_transition}
\end{figure}

The results are shown in Fig.~\ref{fig:secondary_results}, where panel (a) displays the optimal efficiency as function of the fixed output cooling power, while panel (b) displays the optimal efficiency as function of the electrochemical potentials that are required to achieve the optimal efficiency at different fixed cooling powers. Panel (c) shows the relation between output cooling power and electrochemical potential optimizing the efficiency.

The first striking observation is that the optimal efficiency obtained with the nonthermal dip-peak distribution and the one obtained for the equivalent thermal distribution are almost identical. The fact that there is hardly any difference in the results for efficiency and cooling power, as shown in Fig.~\ref{fig:secondary_results}(a)-(c), and the fact that there is hardly any difference in the corresponding transmission functions, as shown in Fig.~\ref{fig:secondary_transition}, for most of the parameters, relies on the almost negligible difference between the temperature and potential of the reference thermal distribution $f_\mathrm{env}$, see Eq.~\eqref{eq:dip_peak_distribution}, and the effective temperature and potential associated to the full dip-peak distribution. 

There is however one crucial difference, resulting from the nonmonotonicity of the nonthermal distribution. In Fig.~\ref{fig:secondary_results}(b,c) a sharp, small feature occurs at $\mu\approx1.5\: k_\mathrm{B}T$, corresponding to a feature at low cooling power in panel~(a). When moving $\mu$ across this value, the cold distribution crosses the peak in the nonthermal distribution, thereby revealing its nonthermal features.
The occurrence of the nonmonotonicities implies that there can be co-existing solutions which fulfill $\lambda = (\partial I^y / \partial \xi |_\mathcal{D})/(\partial I^x / \partial \xi|_\mathcal{D})$ at different $\mu$, but with the same $I^x_\mathrm{fix}$, as seen in Fig.~\ref{fig:secondary_results}(a). These solutions appear since there can be several local extrema of the function $\eta(\mu)$ with fixed output current and optimized transmission function. Figure~\ref{fig:secondary_transition} showcases how a co-existing solution appears as the cold distribution ''passes over'' the peak in the nonthermal distribution, as the transmission window is wider in the middle panel the third row. 

If we compare this sharp, small feature occurring at $I^{\Sigma, \mathrm{L}}\simeq0.015 \:k_\mathrm{B}^2T/h$ with the results in Fig.~\ref{fig:nonthermal_results}(j) obtained \textit{without} optimizing with respect to $\mu$, we note that even the lowest dip value at at $I^{\Sigma, \mathrm{L}}\simeq0.015 \:k_\mathrm{B}^2T/h$ in Fig.~\ref{fig:secondary_results}(a)  has a higher efficiency ($\eta>0.75$) than the one at fixed $\mu=0$ ($\eta\gtrsim0.5$) in Fig.~\ref{fig:nonthermal_results}(j). 

This demonstrates that an optimization with respect to a second parameter can be beneficial, even if there are several solutions yielding different performance. It should however be pointed out that when exploiting nonthermal distributions which deviate more strongly from a thermal one, these co-existing solutions can have a stronger impact when employing the optimization method of Sec.~\ref{sec:opt_two} and should therefore be carefully considered.

\subsection{Approaching boxcar transmission with Lorentzians}\label{sec:lorentzian_transmission}
We have until here found ideal realizations of the transmission functions---namely series of sharp boxcar-shaped transmissions---that optimize the performance of the cooling. These results are important as they set \textit{bounds} on how good the cooling power, the efficiency at fixed cooling power,  or the precision at fixed cooling power can get. 
Realistically, the transmission function in a quantum system is however not expected to be of ideal boxcar shape---fine tuning of for example a large amount of quantum dots coupled in series or similar band engineering would be required~\cite{Whitney2014Apr,Whitney2015}. \reply{The question arises hence how the performance will be impacted if the transmission differs from this ideal case. To achieve boxcar-\textit{like} transmissions one could for example employ quantum dot chains~\cite{Cha2020Oct,Ehrlich2021} together with dedicated tuning of the dot properties~\cite{Kalantre2019Jan}, or exploit metal-molecule-metal junctions where exchanging anchoring groups of the central molecule shapes the electronic transmission~\cite{Tsukamoto2012}. }

\reply{Here, instead of analysing specific, complex realizations potentially approaching a boxcar transmission, we address the example of a transmission function that is also limited in energy but standardly achieved in experiments.  Namely, we analyze a Lorentzian transmission} 
 
\begin{figure}[tb]
    \centering
    \includegraphics[width=\linewidth]{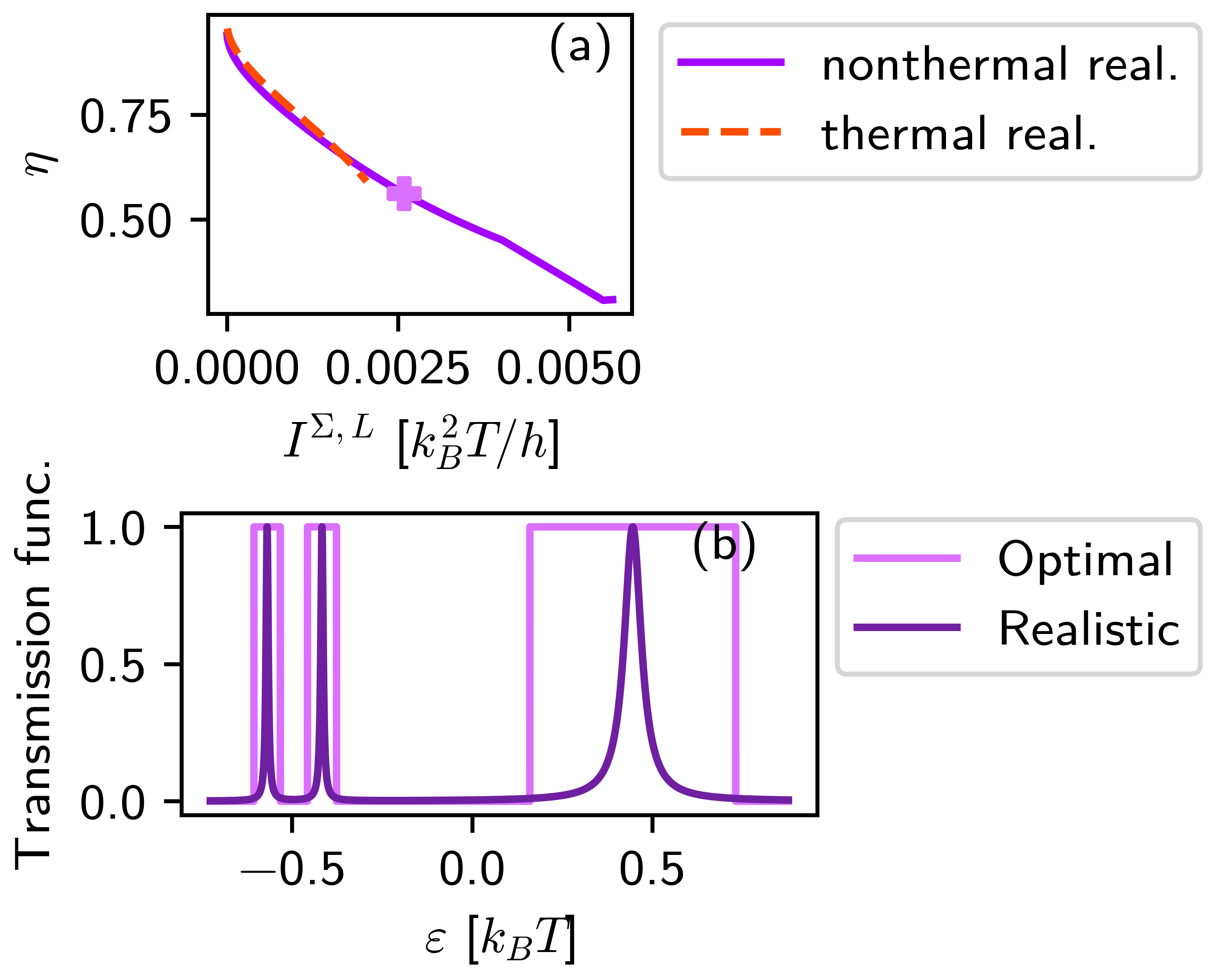}
    \caption{ (a) Efficiency at finite output power \reply{for transmission functions consisting of a series of Lorentzians as exemplified in panel (b).} Results for the nonthermal distribution as well as its thermal equivalent are shown in purple and red. (b) Transmission function consisting of three boxcars optimizing efficiency at \reply{a fixed output power, indicated by the purple plus in panel (a),} for the distribution function of Eq.~\eqref{eq:dip_peak_distribution} (light purple) and series of Lorentzians approximating the ideal case (dark purple). }
    \label{fig:real_transf}
\end{figure}
 
\begin{equation}\label{eq:DLor}
    \mathcal{D}_{\mathrm{Lor},j}(\En) = \mathcal{D}_{0,j}\frac{\Gamma_j^2}{(\En - E_j)^2 + \Gamma_j^2},
\end{equation}
with maximum transmission $\mathcal{D}_{\mathrm{Lor},j}$ found at $\varepsilon=E_j$ and with half width at half maximum $\Gamma_j$.
\reply{Indeed, such} Lorentzian-shaped transmissions are realized in many experimental settings where resonances occur~\cite{Ihn2009Dec}, such as in low-temperature experiments with gate-tunable quantum dots~\cite{Kouwenhoven2001Jun}, in molecular electronics even at room temperature~\cite{Xin2019Mar,Gemma2023Jun}, or in electronic Fabry-Perot interferometers~\cite{Deprez2021May,Camino2007Oct}.
The full transmission of this realistic example \correct{has} the form of a series of resonances
\begin{equation}
    \mathcal{D}_\mathrm{real}(\En) = \sum_j \mathcal{D}_{\mathrm{Lor},j}(\En)\ .
\end{equation}
 We here choose to adapt this series of resonances to the ideal series of boxcar-shaped transmission windows by placing the center of the resonances at the center of the boxcars,  constituting the optimal transmission $\mathcal{D}_\mathrm{opt}(\En)$.\footnote{This should not be confused with an optimization of the efficiency under the constraint that the transmission is a series of Lorentzians, which is not the goal here.} Then, we choose to fix the widths of the Lorentzian transmissions to $0.1 \times$ the width of a given boxcar-shaped part of the optimal transmission. An example for this is shown in Fig.~\ref{fig:real_transf}(b). In this way most of the Lorentzian is contained inside the boxcar transmission that would constitute the ideal case. 

The efficiencies for the Lorentzian transmission functions at given cooling power for the dip-peak distribution and for its thermal equivalent are shown in Fig.~\ref{fig:real_transf}(a) in purple and red respectively. This can be compared with Fig.~\ref{fig:nonthermal_results}(j) for the optimal transmission functions. As expected, the bound on the efficiency, given by the case with an optimized boxcar series, sets an upper limit on the efficiency reached with a series of Lorentzians. Both for the thermal and for the nonthermal distributions this bound is not reached for any finite value of fixed $I^{\Sigma,\mathrm{L}}$. The reason for this is that the algebraic decays of the Lorentzian function include parts of the distributions which result either in a larger $I^{\Sigma, \mathrm{R}}$ than is necessary or negative contributions to the cooling power. This means that electrons are being partially transmitted at lower characteristic efficiencies than the minimal characteristic efficiency determined by the cooling power output, see Eq.~\eqref{eq:opt_eff_char}. 

Let us now compare the performance of a refrigerator using the nonthermal distribution as a resource when the energy filter is a series of Lorentzians with the one where the resource is thermal, see the fully coloured lines in Fig.~\ref{fig:real_transf}. We first notice that the maximum output power reachable with the nonthermal distribution is still larger than the maximum reachable cooling power with the thermal distribution (as given by the end points of the lines in panel~(a)).  
 However, the nonthermal distribution performs slightly worse than the thermal one for most finite values of $I^{\Sigma, \mathrm{L}}$. Towards the end point of the realistic thermal line, the nonthermal starts to perform slightly better. This clearly shows the advantage that a sharp transmission function would have to exploit nonthermal features compared to the Lorentzian transmissions which smoothen out some of those nonthermal features.

Finally, note that there is a large variety of energy-dependent transmission functions in different nanostructures, \correct{while} we here only analyze one example model. By only using one Lorentzian per boxcar of the ideal transmission, the output power is severely lowered due to the necessity of narrow peaks. However, for more complex systems, like a larger and more accurate combination of quantum dots, the realistic transmission function is expected to approach a boxcar shape more closely. What our cursory investigation conveys is that the optimal transmission function can be used to guide the design of realistic functions: the closer they are to the optimal boxcars, the closer the performance should be to the bound. 

\section{Conclusion}\label{sec:conc}

We have developed a general approach to optimize the performance of energy conversion processes in two-terminal coherent conductors operating in steady-state, focusing on devices fed by \textit{nonthermal} distributions and that can be modeled by scattering theory. 
We therefore extended an approach based on Lagrange multipliers put forward in Refs.~\cite{Whitney2014Apr,Whitney2015} to generic performance quantifiers---such as general efficiencies, precision, and trade-off relations---and to the nonthermal case.

We identified the optimal transmission function of the conductor for a given nonthermal distribution and a given performance quantifier as a series of boxcar-shaped transmission windows, which can be experimentally approached, for example by a collection of well-tuned quantum dots. The position and width of the optimal transmission windows are determined by crossing points between spectral properties and characteristic functions determined by imposed constraints (here a fixed output current). This optimal case sets a bound on the performance that can be achieved in an energy-conversion process exploiting a nonthermal resource. 

Nonthermal distributions can arise in small-scale steady-state conductors due to multiple reasons. Our work shows the potential advantage of such distributions for energy-conversion processes.
We exemplified our approach at the example of cooling a cold thermal contact using two realistic nonthermal distributions, as they could arise from light irradiation or from two competing environments. We found that the optimum performance is at least as good but often significantly better than the one that can be reached by an equivalent thermal distribution, which would provide the same amount of energy or particle flow (mimicked by B\"uttiker probes).
 
 Notably, the optimal transmission functions are \textit{different} for the nonthermal and for thermal resources. In fact, we find that the \textit{nonmonotonicity} in the energy spectrum of nonthermal distribution leads to more fine-grained energy transmission windows for the transmission function which optimally exploit the resource. Therefore, if the nonthermal nature is not taken into account when designing nanoelectronic devices, the performance may suffer.

While the focus of this paper is on nonthermal resources, we underline that the optimization methods are \textit{general}, and expand in generality and versatility the previous results for optimizing energy conversion. This is especially achieved by extending the method to the optimization of general trade-off relations, which---if desirable---could be further generalized to more complex functions of performance quantifiers. 

We suggest that the optimization methods we have devised could provide guidelines for improving and benchmarking performance in realistic nanoelectronic devices by setting performance bounds. Furthermore, we anticipate that by designing energy conversion processes according to the optimal transmission function, performance could be improved. We also urge for more investigations into occurrences of nonthermal distributions in nanoelectronic devices, since these can demonstrably have a positive impact on performance. 

\acknowledgments
We thank Gabriel Landi for helpful discussions. We thank Bruno Bertin-Johannet, Juliette Monsel and Didrik Palmqvist for providing us with useful feedback on our manuscript. Funding from the Knut and Alice Wallenberg Foundation via the Fellowship program and  from the European Research Council (ERC) under the European Union’s Horizon Europe research and innovation program (101088169/NanoRecycle) is gratefully acknowledged.

\section*{Data availability}
The data that support the findings are openly available \cite{zenodo}.

\appendix

\section{Crossing point illustrations}\label{app:crossings}
The optimization results in Sec.~\ref{sec:opt} show that the optimal transmission function is a series of boxcars regardless of performance quantifier. Naturally, the locations of the boxcars in the energy spectrum depend on the specific quantifier and are determined by full transmission conditions, namely the functions inside the Heaviside functions in Eqs.~\eqref{eq:opt_eff_transmission},~\eqref{eq:opt_noise_transmission}, and~\eqref{eq:opt_transmission_product}. In this Appendix, these conditions are investigated further.

\begin{figure}[h!]
    \centering
    \includegraphics[width=\linewidth]{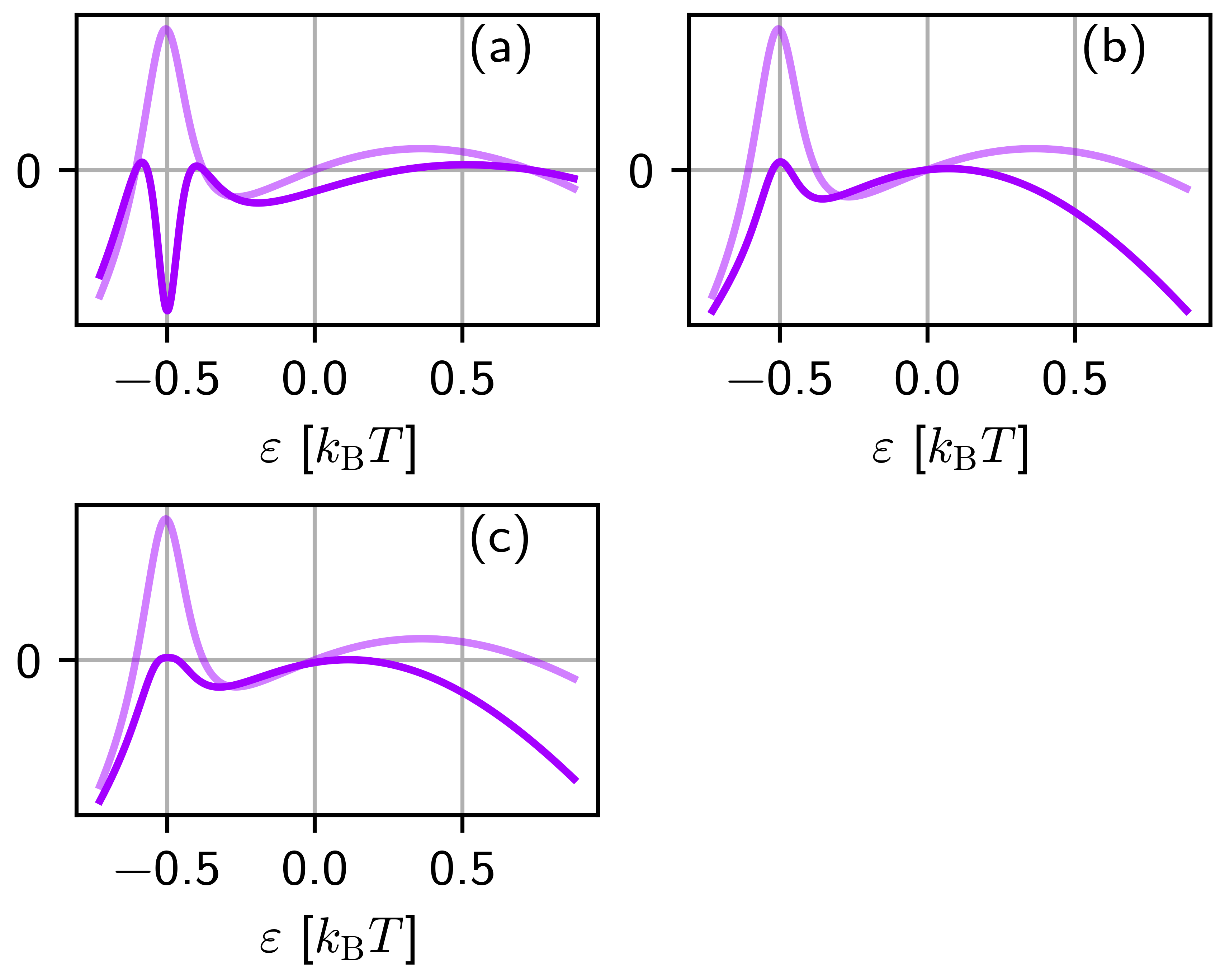}
    \caption{Illustrative graphs of the full transmission conditions for fixed current $I^{\Sigma, \mathrm{L}} = 0.008 \: \kB T$ and minimized current $I^{\Sigma, \mathrm{R}}$ between the dip-peak and cold distributions. In all panels, the functions are given in arbitrary units. Figures~\ref{fig:cross_point_all}(a)-(c) depict the argument of the Heaviside functions in Eqs.~\eqref{eq:opt_eff_transmission},~\eqref{eq:opt_noise_transmission}, and~\eqref{eq:opt_transmission_product} respectively in strong purple. The weak purple lines are all plots of the argument of the Heaviside function in Eq.~\eqref{eq:opt_max_transmission}, which defines the cooling windows. For all functions, values above zero mean that the transmission function is set to 1 for those energies and vice versa.}
    \label{fig:cross_point_all}
\end{figure}

 In Figs.~\ref{fig:crossing_eff} and~\ref{fig:char_eff} the condition for optimal efficiency was broken down into its components and rewritten in terms of the characteristic efficiency, see Eq.~\eqref{eq:char_eff}, the latter requiring that the currents are well-behaved, as in fulfilling the Second Law. While these analyses are useful for physical understanding, it can be beneficial to consider the conditions in full to get a quick overview of where the transmission windows are located and how they change with the output current. Figure \ref{fig:cross_point_all} shows the functions that determine the condition for full transmission for (a) optimized efficiency, (b) optimized precision and (c) optimized trade-off. Thus, panels~(a), (b), and (c) of Fig.~\ref{fig:cross_point_all} show examples of the functions inside the Heaviside functions in Eqs.~\eqref{eq:opt_eff_transmission},~\eqref{eq:opt_noise_transmission}, and~\eqref{eq:opt_transmission_product} respectively, in comparison with the function for the maximal output, Eq.~\eqref{eq:opt_max_transmission}, plotted with a fainter line. Let us call these ``condition functions'' for brevity.
 All functions are in arbitrary units. For any graph, a value larger than zero means that the corresponding transmission function is set to 1, which in particular means that only the sign of the function has a meaning.  From these panels, one can directly read out where to place the boxcar transmissions.
 
As a next step, consider $\lambda$ in the condition functions. Once the expressions for the condition functions, and thus the respective optimal transmission functions, are found, we can consider the output current as a function of $\lambda$, $I^x = I^x(\lambda)$. By regarding $\lambda$ as a variable, the spectrum of possible $I^x$ is found by evaluating the current integral with the optimal transmission functions. The minimum and maximum currents are found in the limits of $\lambda \in [0,\infty)$. As $\lambda \to \infty$, the output current reaches the maximum possible current. This means that the optimal transmission function reaches the transmission function for maximum output current.  Furthermore, the condition functions for the optimized quantities must converge to the one for maximum cooling power. How close the functions are to the maximum condition function depends on $\lambda \in [0,\infty)$, where larger values mean that the optimal transmission functions encompass larger areas of the cooling window(s).  

As such, one can read out from the panels in Fig.~\ref{fig:cross_point_all} how the function should change with increasing $\lambda$. For example, in panel (a) the crossing point at around $\En = 0.25$ will move towards $\En = 0$. Meanwhile, the dip at $\En = -0.5$ will shift upward and invert, in order to converge to the peak of the maximum condition function. In particular, this means that for some $\lambda$ there will be one boxcar transmission instead of two in the cooling window around $\En = -0.5$. This ``merging'' of boxcar functions is likely why the features in Fig.~\ref{fig:secondary_results} appear.
\begin{figure*}[htb]
    \subfloat{
        \includegraphics[width = 0.45\linewidth]{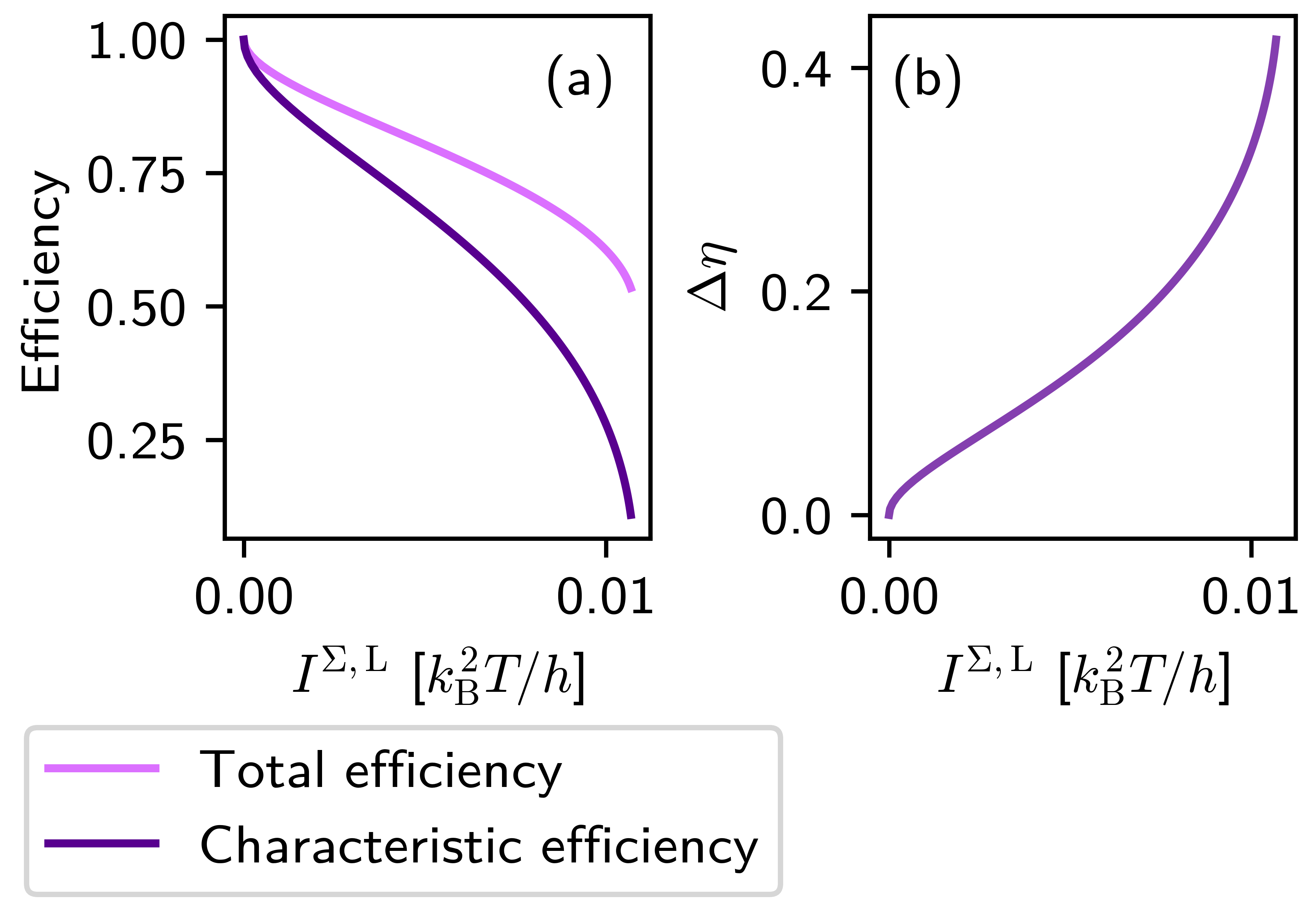}}
    \hfill
    \subfloat {
        \includegraphics[width = 0.45\linewidth]{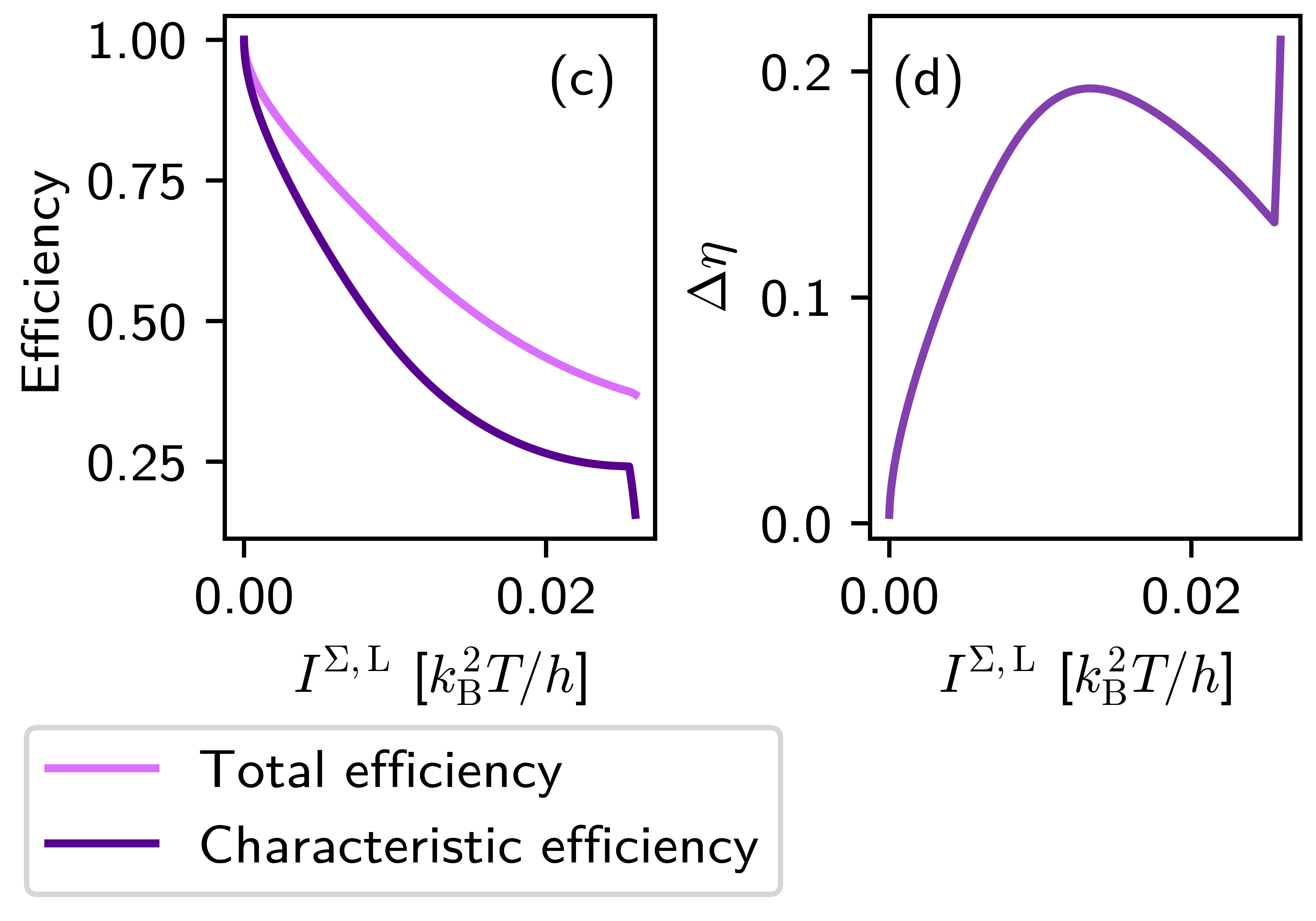}}
    \caption{Comparison between characteristic and total efficiency for the mixed distribution (a,b) and the dip-peak distribution (c,d). Panels (a) and (c) show the efficiency functions, where the light purple line for total efficiency corresponds exactly to the optimized efficiencies in Fig.~\ref{fig:nonthermal_results}(b) and (j). The dark purple line is the corresponding minimum characteristic efficiency to the optimal total efficiencies, see Eq.~\eqref{eq:opt_eff_char}. Panels (b) and (d) show the difference between the total and characteristic efficiencies. Note that graphs are quite different in the two panels, suggesting that there is no universal function between the characteristic and total efficiency.}
    \label{fig:char_eff_comparison}
\end{figure*}

While the full condition can be insightful, it is given further meaning by rewriting in terms of defined characteristic functions. In particular, one can discuss the characteristic efficiency defined in Eq.~\eqref{eq:char_eff} and used to express the optimal transmission function for efficiency in Eq.~\eqref{eq:opt_eff_char}. The characteristic efficiency is interpreted as the efficiency for a single energy channel, or the efficiency which an individual electron ``carries''. This begs the question of whether there is a direct connection between the characteristic and the total efficiency, the latter being defined in Eq.~\eqref{eq:def_efficiency}. This would be quite convenient and might even be said to be expected. However, as Fig.~\ref{fig:char_eff_comparison} shows, there is no clear algebraic relationship between the two. Panels (a) and (c) (mixed and dip-peak) depict the total efficiencies for optimized transmission functions over the possible range of $I^{\Sigma, \mathrm{L}}$ compared with the corresponding minimum characteristic efficiencies. The right-hand panels (b) and (d) (mixed and dip-peak) show the absolute difference between the two. Clearly, there is no simple linear relationship between the two quantities. Furthermore, the difference functions are vastly different between the mixed and dip-peak cases. Note also the discontinuity near the end of the spectrum for the dip-peak case in panels (c) and (d). This appears due to two boxcars joining into one, when the valley in Fig.~\ref{fig:cross_point_all}(a) crosses completely over zero. Together with the inconsistent relationship between the total and minimal characteristic efficiencies, the conclusion is that there is no obvious and universal analytical function $\eta = \eta(\eta^\mathrm{char}_\mathrm{min})$.

What can be surmised, however, is that the total efficiency is consistently higher than the minimal characteristic efficiency. This is not a surprising fact, as most of the contributions to the total efficiency are made by electrons carrying larger characteristic efficiencies than the minimum by design. Then the minimal characteristic efficiency can be seen as a lower bound on the total efficiency. 

\section{Noise minimization based on thermal-like contribution only}\label{app:only_thermal}
In Sec.~\ref{sec:gen_noise}, we have argued that for the minimization of noise, it is sufficient to consider the thermal noise, since for full or completely suppressed transmission, namely for $D(\epsilon)\in\{0,1\}$, the shot noise (which would otherwise always be detrimental) disappears. This can also be argued for by considering the energy spectrum again.

We start by considering general properties of the functions to be optimized. We therefore first recall the definition of a concave function, $f: \mathbb{R} \rightarrow \mathbb{R}$, which is
\begin{equation}
    f(\alpha x  + (1-\alpha) y) \geq \alpha f(x) + (1-\alpha)f(y), \alpha \in [0,1]
\end{equation}
for $x,y$ two values on the real axis. This means that the value of the function for any point between $x$ and $y$ is larger than or equal to the values at the point with the lowest value. It is straightforward to show that the sum of two concave functions is also concave, leading to the notion of concavity in the integral. 

We first set up a constraint equation for the discretized noise and corresponding current, 
\begin{equation} \label{eq:discrete_setup_noise}
    S^x = \sum_\gamma S^x_\gamma(d_\gamma) +\lambda \big(I^x_\mathrm{fix} - \sum_\gamma I^x_\gamma (d_\gamma)\big),
\end{equation}
As before, we consider one of the energy slices 
\begin{eqnarray}
    &x^2_\gamma d_\gamma \Big[(1-d_\gamma)[f_\gamma - g_\gamma]^2 +\\
    &g_\gamma (1-g_\gamma) +f_\gamma (1-f_\gamma) \Big]\delta \En   - \lambda x_\gamma d_\gamma (f_\gamma - g_\gamma) \delta \En\nonumber\label{eq:complicated_noise}
\end{eqnarray}
where $d_\gamma$ is the height of the transmission at $\varepsilon_\gamma$. An equivalent, more easily digestible function is
\begin{equation}
    h(d_\gamma) = a_\gamma d_\gamma  + b_\gamma d_\gamma(1-d_\gamma) = d_\gamma(a_\gamma + b_\gamma) - b_\gamma d_\gamma^2
\end{equation}
where we introduced the constants $a_\gamma$ and $b_\gamma$ for a certain slice $\gamma$ comparing to the prefactors of $d_\gamma$ and $d_\gamma(1-d_\gamma)$ in Eq.~\eqref{eq:complicated_noise}. This function is clearly a concave hyperbolic function, due to the negative square term. When integrating, one sums over all the slices $\gamma$ and takes the limit $\delta \epsilon \rightarrow0$. Since the sum is a concave function, we can conclude that the integral $S^x$ is a concave functional of $\mathcal{D}(\varepsilon)$ which implies that there is no local minimum and the global minimum is on the bounds of the function domain, meaning that the transmission must be either zero or one at every energy for the global minimum solution. A more rigorous argument for this can be found in~\cite{Landi2025}.

\section{Classification of spectral currents} \label{app:spectral}
In this Appendix, we expand on the reasoning as to why Eq.~\eqref{eq:opt_transf_before_char_eff} follows from Eq.~\eqref{eq:two_contributions_D} and explain why $\lambda >0$ is picked. 

Recall that the condition that determines the optimal transmission function for efficiency is expressed through the spectral currents as $i^y(\En) < \lambda i^x(\En)$. Since these two spectral currents are directly compared to form the optimal transmission function, it is vital to understand their physical meaning. In Table~\ref{tab:spectral_classification} the four possible operating regimes are classified according to the signs of the spectral currents. Note that the total currents, $I^x$ and $I^y$, are always fixed to be positive in our performance analysis, but that the current integrals can still span over energies where the spectral currents are negative. 

Only two of the four regimes presented in Table~\ref{tab:spectral_classification} are in principle supposed to be relevant for the optimization: the regular and the reversed operation. The other two are either forbidden or not beneficial. The forbidden region refers to a a set of conditions under which both a resource and an output are generated simultaneously, which should be impossible assuming the currents are well-defined and comply with the second law of thermodynamics. If such a region were to exist, the transmission function could be defined exclusively within it, leading to an unphysical system. For instance, consider a case where the resource current corresponds to the entropy production in one contact and takes a negative value, while the output current represents the entropy reduction in another contact (i.e., negative entropy production). The presence of a forbidden region would then imply the possibility of constructing a device with negative total entropy production, as entropy would be reduced in both contacts. 

The non-beneficial region, on the other hand, can exist. Using the previous example, the entropy would increase in both contacts, which is physically allowed. However, this regime provides no possible benefit to the performance and is hence excluded from a device operation that is supposed to have optimal efficiency.

\begin{table}[h!]
    \centering
    \begin{tabular}{c|c|c}
         & $i^x(\En) > 0$ & $i^x(\En) < 0$\\ \hline
        \multirow{3}{*}{$i^y(\En) > 0$} & resource spent & resource spent\\
        & output generated & output spent \\
        & \green{Regular op.} & \red{No benefit}\\ \hline
        \multirow{3}{*}{$i^y(\En) < 0$} & resource generated &  resource generated\\
        & output generated & output spent \\ 
        & \red{Forbidden} & \red{Reversed op.} \\ 
    \end{tabular}
    \caption{Classification of operating regimes according to the signs of the resource and output spectral currents. }
    \label{tab:spectral_classification}
\end{table}
From the above discussion, we conclude that the parameter $\lambda$, introduced as a Lagrange multiplier, must be positive in order for the output current to be positive. This requirement follows from the sign constraints already imposed on both the total currents and the spectral currents. As shown in Eq.~\eqref{eq:spectral_inequality}, $\lambda$ appears as a prefactor to the spectral current associated with the output, $i^x(\En)$. If $\lambda < 0$, the transmission function would equal 1 in regions where $i^y(\En) < 0$ and $i^x(\En) > 0$, or where $i^y(\En) > 0$ and $i^x(\En) < 0$. These two situations correspond to forbidden or nonbeneficial regimes, meaning the resulting transmission function would not produce a positive output current. Therefore, one can safely assume that $\lambda > 0$ without constraining the positive solutions to the output current. 

The classification is also used to exclude the second line in Eq.~\eqref{eq:two_contributions_D}, which should always evaluate to zero. The first Heaviside function in the first line points to the first row of Table~\ref{tab:spectral_classification}. To fulfill the condition $i^x(\En)/i^y(\En) > 1/\lambda$, $i^x(\En)$ has to be positive since $i^y(\En)>0$, by virtue of $\lambda > 0$. Thus, only the regular operation is included and the nonbeneficial regime is naturally excluded.

By contrast, the second line contributes only when $i^y(\En) < 0$, meaning the resource is \textit{generated} in that energy window (while the integrated resource current $I^y$ remains constrained to be positive), as indicated in the second row of Table~\ref{tab:spectral_classification}. In principle, both regimes can satisfy $\frac{1}{\lambda} - \frac{i^x(\En)}{i^y(\En)} > 0$ for $\lambda > 0$. The forbidden regime would always satisfy it, but—by earlier arguments—cannot occur. The reversed-operation regime, however, does exist and can not be per se excluded. One might imagine partially including reversed operation to generate the resource, which could then be consumed during regular operation to produce output. Physically, however, it is impossible to increase the efficiency by adding this part of the spectrum; such a process would amount to constructing a perpetual-motion device for the following reason. The efficiency could only be increased by adding this part of the spectrum, if there were regions in reversed operation where the resource gained exceeds the output spent, violating the first or second law of thermodynamics, when expressed in common units. For example, in the case of entropy currents this would correspond to a net reduction of entropy, violating the second law. Consequently, including reversed-operation energy windows cannot enhance the device’s efficiency.

\section{Numerical procedure}\label{app:numerical}
Several results presented in Sec.~\ref{sec:optimal_fridge} (Figs.~\ref{fig:nonthermal_results} and~\ref{fig:secondary_results} in particular) are calculated fully numerically. This Appendix summarizes the numerical procedure that was used to obtain these results. As per the data availability statement, the code and data used are all freely available and further details can be found there~\cite{zenodo}. 

The code for producing the data is written in Python and the numerical method relies heavily on the SciPy module~\cite{2020SciPy-NMeth} for equation solvers and integrations. Principally, the challenge of applying the methods in Sec.~\ref{sec:opt} lies in the Heaviside functions; because large parts of the energy spectrum may be zero, especially for low fixed currents, standard integration methods can produce erroneous results, usually returning zero. Instead, the integration is divided into several areas, defined by the roots of the condition inside the Heaviside functions. A standard integration is made over each of these areas and added together to form the calculated current or noise. This ensures accurate results for any optimal transmission function, especially correcting for cases like Fig.~\ref{fig:nonthermal_results}(n), where the boxcars are narrow and disconnected. The roots calculation does add some computational complexity, but the impact is small in our implementation despite the brute-force minimization over the energy spectrum. 

The analytical optimal transmission functions found in Sec.~\ref{sec:opt} depend on a Lagrange multiplier $\lambda$ to fix the current $I^x_\mathrm{fix}$. If there is therefore a picked $I^x_\mathrm{fix}$, $\lambda$ can be solved for numerically by inserting the expression for $\mathcal{D}(\En)$ into the integral equation for the current, Eq.~\eqref{eq:generic_current}. When producing data over the fixed current spectrum, like in Fig.~\ref{fig:nonthermal_results}, one can instead exploit that $\lambda$ is a function of the fixed current, $\lambda = \lambda(I^x)$. Practically, for the three optimized quantifiers in Fig.~\ref{fig:nonthermal_results} the current integrals were calculated over a range of $\lambda$, between the minimum (1 for efficiency optimization, 0 for the others) and a large enough value to approach the maximum output current; recall that $\lambda$ is unbounded from above. This simplifies and quickens the procedure, as there is no need for any equation solver, in case of efficiency and precision optimization.

For the trade-off optimization, interchangeably called product optimization in the code, there is a need for equation solving. This is because there are three unknown variables and only one of them is effectively eliminated. For each $\lambda$ in the range, the noise and resource current must be solved for self-consistently. 

There is a similar simplification made for the results with second variable optimization, see Fig.~\ref{fig:secondary_results}, but it is still necessary to solve an equation. As the problem is set up in Sec.~\ref{sec:opt_two}, one solves for a fixed current by finding the $\xi$ which yields $\lambda = (\partial I^y / \partial \xi |_\mathcal{D})/(\partial I^x / \partial \xi|_\mathcal{D})$, with $\mathcal{D}(\En)$ determined by the optimal transmission function for efficiency, in this case. To find the results in Fig.~\ref{fig:secondary_results}, we essentially do the opposite: for a range of $\mu$s, find $\lambda$ such that $\lambda = (\partial I^y / \partial \xi |_\mathcal{D})/(\partial I^x / \partial \xi|_\mathcal{D})$. This simplifies the procedure, since it is easier to control the range of $\mu$s than the range of $I^x$, and allows for the co-existing solutions found in Fig.~\ref{fig:secondary_results} seen in the dips of the nonthermal curve. 

For all equation solvers used all data is saved. When the data is presented in figures in the paper, data with too large errors are excluded. These are points where the equation solver did not return solution within reasonable precision. This precision is set to $10^{-7}$, although many errors are orders of magnitude smaller. Relevant error data is available through the data statement.  

\bibliography{refs}

\end{document}